\documentclass[12pt]{interact}
\usepackage[T1]{fontenc}
\usepackage{lmodern}

\usepackage{epstopdf}
\usepackage[caption=false]{subfig}
\usepackage{color}

\usepackage[numbers,sort&compress]{natbib}
\bibpunct[, ]{[}{]}{,}{n}{,}{,}

\theoremstyle{plain}

\theoremstyle{definition}

\theoremstyle{remark}

\begin{document}
\title{Sum rules for x-ray circular and linear dichroism based on complete magnetic multipole basis}

\author{
\name{Y. Yamasaki\textsuperscript{a,b,c}, Y. Ishii\textsuperscript{a}
and 
N. Sasabe\textsuperscript{a}}
\thanks{CONTACT Yuichi Yamasaki Email: YAMASAKI.Yuichi@nims.go.jp} 
\affil{\textsuperscript{a}Center for Basic Research on Materials, National Institute for Materials Science (NIMS), Tsukuba 305-0047, Japan; 
\textsuperscript{b}International Center for Synchrotron Radiation Innovation Smart, Tohoku University, Sendai 980-8577, Japan; 
\textsuperscript{c}Center for Emergent Matter Science (CEMS), RIKEN, Wako 351-0198, Japan
}
}

\maketitle

\begin{abstract}
X-ray magnetic circular dichroism (XMCD) and X-ray magnetic linear dichroism (XMLD) are powerful spectroscopic techniques for probing magnetic properties in solids. 
In this study, we revisit the XMCD and XMLD sum rules within a complete magnetic multipole basis that incorporates both spinless and spinful multipoles. 
We demonstrate that these multipoles can be clearly distinguished and individually detected through the sum-rule formalism. 
Within this framework, the anisotropic magnetic dipole term is naturally derived in XMCD, offering a microscopic origin for ferromagnetic-like behavior in antiferromagnets. 
Furthermore, we derive the sum rules for out-of-plane and in-plane XMLD regarding electric quadrupole contributions defined based on the complete multipole basis.
Our theoretical approach provides a unified, symmetry-consistent framework for analyzing dichroic signals in various magnetic materials. 
These findings deepen the understanding of XMCD and XMLD and open pathways to exploring complex magnetic structures and spin-orbit coupling effects in emergent magnetic materials.
\end{abstract}


\section{Introduction}
\subsection{X-ray absorption spectroscopy}
X-ray absorption spectroscopy (XAS) is a powerful tool for investigating the electronic and magnetic properties of materials \cite{Groot2008}. 
Among its various techniques, X-ray Magnetic Circular Dichroism (XMCD) and X-ray Magnetic Linear Dichroism (XMLD) have been extensively applied in synchrotron-based experiments to investigate element-specific information on magnetism \cite{Schuetz1987,Stohr1999, Alders1998, vanDerLaan2014}.
XMCD is a phenomenon where the absorption of circularly polarized X-rays differs depending on the relative orientation of the photon helicity and the magnetization. 
This effect arises from the spin-orbit interaction in the core-level states, leading to different transition probabilities for left- and right-circularly polarized light.
XMCD is widely used to determine element-specific spin and orbital magnetic moments via sum rules \cite{Thole1992, Carra1993, Chen1995, TanakaJo1996JPSJ}.
It has been extensively applied to transition metal and rare-earth compounds to study magnetic ordering, local electronic structure, and hybridization effects.
XMCD has played a crucial role in understanding magnetic thin films, multilayers, and nanostructures, especially in spintronics and permanent magnets
\cite{vanDerLaan2014, Chong2014PRL, Zhang2015PRB, Miwa2015APL, Chang20219ACSAPP}.
Due to its sensitivity to both spin and orbital contributions to magnetism, XMCD enables separating these components \cite{ishii2024microscopic}, which is essential for studying phenomena such as spin-orbit coupling and anisotropy in magnetic materials.

In contrast to XMCD, XMLD refers to the difference in X-ray absorption between two orthogonal linear polarizations in a magnetically ordered system \cite{ChenXMLD1997,scholl2000observation}. 
XMLD is primarily sensitive to the anisotropy of the local electronic structure and provides information about orbital occupation and magnetic ordering.
Unlike XMCD, which directly probes net magnetization, XMLD is particularly useful for investigating antiferromagnetic and non-collinear magnetic structures.
XMLD arises from the anisotropic valence state due to the spin-orbit and exchange interactions, leading to an anisotropic absorption cross-section. 
This makes XMLD a valuable tool for studying magnetocrystalline anisotropy, spin reorientation transitions, and antiferromagnetic domain structures in materials such as transition metal oxides and rare-earth compounds.

\subsection{Magnetic dipole order in Antiferromagnet}
Recent studies in magnetism have revealed that phenomena traditionally associated with ferromagnets, such as the anomalous Hall effect (AHE), the magneto-optical Kerr effect (MOKE), and XMCD can emerge even in antiferromagnetic materials, provided that the magnetic ordering symmetry allows for magnetic dipole components. 
This behavior has been observed in both non-collinear antiferromagnets, such as Mn$_3$Sn \cite{Nakatsuji2015,Mn3Sn_2018_MOKE_higo}, and collinear antiferromagnets recently classified as altermagnets, including manganese oxides \cite{Solovyev1997,Noda2016_PhysChemChemPhys,Okugawa_AM2018}, organic antiferromagnetic \cite{Naka2019NatCom}, RuO$_2$ \cite{Ahn2019PRB,Smejkal2020_SciAdv, Smejkal2022, Smejkal2023_NatRevMater} and MnTe \cite{Hariki2024PRL, MnTe2024Nature}.
In these systems, although the net magnetization vanishes in real space, the presence of symmetry-allowed magnetic dipole moments, $i.e.$ specifically the anisotropic magnetic dipole term ($t_z$), enables ferromagnetic-like responses such as AHE and XMCD to appear \cite{Yamasaki2020,Sasabe_PRL_Mn3Sn, XMCD_Kimata,sakamoto2021observation, sakamoto2024bulk}. 
This underscores that the key condition for the emergence of these effects is not the specific spin configuration (collinear or non-collinear), but the magnetic dipole symmetry permitted by the crystal and magnetic structure.
These observations collectively emphasize that optical responses such as XMCD do not directly probe the net magnetization, but are determined by the magnetic multipole symmetry inherent to the system. 
Therefore, to properly interpret the spectroscopic signals, including XAS and XMLD, it is essential to establish the corresponding sum rules within a complete magnetic multipole framework, ensuring all contributions from hidden multipole moments.

\subsection{Complete multipole basis}
Multipole expansions systematically describe various physical properties, such as charge and magnetic distributions, in condensed matter systems. 
The conventional classification distinguishes between electric multipoles $\bm{Q}$ (arising from charge distributions) and magnetic multipoles $\bm{M}$ (associated with current and spin distributions). 
However, recent developments by Hayami and Kusunose $et~al.,$ have established a complete magnetic multipole basis that extends this classification to include electric toroidal multipoles $\bm{G}$ and magnetic toroidal multipoles $\bm{T}$ \cite{kusunose2020complete,hayami2024unified}. 
The multipole representation of four-type multipoles enables a comprehensive understanding of various electronic properties and physical phenomena observed in materials.

The general form of a multipole operator in spinless Hilberlt space is expressed as $\hat{X}_{lm}^{(\text{orb})}$ where \( l \) and \( m \) denote the quantum numbers of the orbital angular momentum, and $X$ represents the type of multipole ($\bm{X}=\bm{Q},\bm{M},\bm{T},\bm{G}$).
The multipole operator in the spinful space is obtained by angular momentum coupling between the spinless multipole and the spin angular momentum \cite{kusunose2020complete, hayami2024unified}, which is given by
\begin{equation}\label{multipole}
    \hat{X}_{lm}^{(s,k)}\equiv i^{s+k}\sum_{n=-s}^s C_{l+k,m-n;sn}^{lm}\hat{X}_{l+k,m-n}^{(\text{orb})}\hat{\sigma}_{sn}
\end{equation}
where $C_{l_1,m_1;l_2,m_2}^{lm}$ is the Clebsh-Gordan coefficient, $s=0,1$ and $-s\le k \le s$ ($k$ is integer).
The spin operator is expressed using the Pauli matrices $\sigma_{sn}$ in the spin space, defined as $\sigma_{00} = \sigma_0$ (the identity matrix), $\sigma_{10} = \sigma_z$, and $\sigma_{1,\pm1} = \mp (\sigma_x \pm i\sigma_y) / \sqrt{2}$.
Magnetic multipoles are classified based on their transformation properties under spatial inversion (\(\mathcal{I}\)) and time-reversal (\(\mathcal{T}\)) symmetry.
In this paper, we focus only on multipoles that possess spatial inversion symmetry (\(\mathcal{I}\)), namely, electric monopoles $Q_{00}$, magnetic dipoles $M_{1m}$, and electric quadrupoles $Q_{2m}$.

Spinless multipoles ($s=0$) are expressed as $\hat{X}_{lm}^{(0,0)} \equiv \hat{X}^{(\text{orb})}_{lm} \sigma_{0}$.
Since $\sigma_0$ is the identity matrix, the type of magnetic multipole basis coincides with that of the orbital multipole basis, namely, $\hat{Q}_{lm}^{(0,0)} = \hat{Q}^{(\text{orb})}_{lm}$ and $\hat{M}_{lm}^{(0,0)} = \hat{M}^{(\text{orb})}_{lm}$.
On the other hand, in the spinful space ($s=1$), since the time-reversal symmetry of spin is odd, the time-reversal symmetry of $\hat{X}_{lm}^{(1,k)}$ should be opposite to that of $\hat{X}^{(\text{orb})}_{lm}$.
In addition, since the spin is the axial vector, a spinful multipole is composed of orbital multipoles with different spatial parity.
For example, the electric multipole ($\hat{X}=\hat{Q}$) contains three spinful multipoles ($s=1$, $k=-1,0,1$),
\begin{align}
    \hat{Q}_{lm}^{(1,0)} &= i\sum_n C_{l,m-n;1n}^{lm} \hat{T}_{l,m-n}^{(orb)}\hat{\sigma}_{1n},\\
    \hat{Q}_{lm}^{(1,\pm 1)} &= i^{1\pm 1}\sum_n C^{lm}_{l\pm 1,m-n;1n} \hat{M}_{l\pm 1,m-n}^{(orb)}\hat{\sigma}_{1n},
\end{align}
suggesting that spinful charge multipoles are generated from the combination of magnetic toroidal moments with the same $l$ ($\hat{T}_l$) and magnetic multipoles with $l$ differing by one ($\hat{M}_{l\pm 1}$).
As well as, the magnetic multipole ($\hat{X}=\hat{M}$) has
\begin{align}
    \hat{M}_{lm}^{(1,0)} &= i\sum_n C^{lm}_{l,m-n;1n} \hat{G}_{l,m-n}^{(orb)}\sigma_{1n},\\
    \hat{M}_{lm}^{(1,\pm 1)} &= i^{1\pm 1}\sum_n C^{lm}_{l\pm 1,m-n;1n} \hat{Q}_{l\pm 1,m-n}^{(orb)}\sigma_{1n},
\end{align}
indicating that spinful charge multipoles are generated from the combination of electronic toroidal moments with the same $l$ ($\hat{G}_l$) and electric charge multipoles with $l$ differing by one ($\hat{Q}_{l\pm 1}$).
The relationship between these complete magnetic multipole bases and the spinless and spinful bases is summarized in Table 1.

The matrix element of $\hat{X}_{lm}^{(s,k)}$ on $|l_v,m_v;\frac{1}{2}m_s \rangle$ basis is given by
\begin{equation}\label{kappamu}
\langle l_vm_v^\prime;\frac{1}{2}m_s^\prime|\hat{X}_{lm}^{(s,k)}|l_vm_v;\frac{1}{2}m_s\rangle = i^{s+k} 
C_{\kappa\mu;sn}^{lm}
C_{l_vm_v;\kappa\mu}^{l_vm_v^\prime}
C_{\frac{1}{2}m_s;sn}^{\frac{1}{2}m_s^\prime}
\frac{\langle l_v||{\hat{X}}_{\kappa}^{\text{(orb)}}||l_v\rangle}{\sqrt{2l_v+1}}
\end{equation}
with $\kappa \equiv l+k$ and $\mu \equiv  m-n$.
$\langle l_v||\hat{X}_{\kappa}^{\text{(orb)}}||l_v\rangle$ is the reduced matrix element, whose explicit expression for $X=Q$ and $M$ are given in Eqs. (8) and (9) in Ref. \cite{kusunose2020complete}.
Using the spherical tensor for multipole for $l=0,1,2,\cdots$ and its $z$-component $m=-l,-l+1,\cdots, l$, $O_{lm} \equiv \sqrt{\frac{4\pi}{2l+1}}r^lY_{lm}(\hat{r})$ with $\hat{r}=\bm{r}/r$ and the spherical harmonics $Y_{lm}$, the electronic and magnetic multipoles are given by $Q^{\text{(orb)}}_{lm} = O_{lm}$ and 
\begin{equation}\label{magnetic_multipoles}
M^{(\mathrm{orb})}_{l,m} = \frac{1}{2} \left[ (\nabla O_{lm}) \cdot \hat{m}_l + \hat{m}_l \cdot (\nabla O_{lm}) \right]    
\end{equation}
with $\hat{\bm{m}_l} = 2\hat{\bm{l}}/(l+1)$, respectively.
For example, $\nabla O_{lm}$ corresponding to the magnetic dipole ($l=1$) and octupole ($l=3$) can be explicitly given in Table \ref{nabla_O_lm}.

\section{Theoretical Framework of X-ray absorption}
In recent decades, the development of sum rules for X-ray absorption spectroscopy has significantly advanced our understanding of the electronic and magnetic properties.
Sum rules link the integrated intensity of XAS and dichroism spectra to ground-state quantities such as spin, orbital magnetic moments, and charge distribution.
These relationships have provided profound insights into magnetic anisotropy, spin-orbit coupling, and electronic correlations.
This work presents sum rules for X-ray absorption based on the complete magnetic multipole bases.

The theoretical framework for X-ray absorption for electric dipole transition is rooted in the Fermi Golden Rule, which relates the absorption coefficient to the transition probability $\mu_\varepsilon(\hbar\omega)$ defined by
\begin{equation}\label{fermi}
    \mu_\varepsilon(\hbar\omega) \equiv \sum_{f,M} \left| \langle \psi_f | \bm{E}\cdot \bm{r} | \psi_g \rangle \right|^2 \delta(E_f - E_g - \hbar\omega), 
\end{equation}
with polarization vector $\bm{E}=E_0\bm{\varepsilon}$ of the incident x-ray polarization and position operator $\bm{r}$.
They can be expanded as the vector product in terms of spherical tensor operators,
\begin{equation}
    \mbox{\boldmath $E$} \cdot \mbox{\boldmath $r$} = E_0 \sum_{M=-1}^{1} (-1)^{M} \varepsilon_{-M} r_{M}
\end{equation}
where $r_{M}$ is the spherical tensors of rank 1, and $\varepsilon_{\pm 1} =\mp(\varepsilon_x\pm i\varepsilon_y)/\sqrt{2}$ and $\varepsilon_0=\varepsilon_z$, corresponding to the circular and linear polarization, respectively.
$E_g$ ($E_f$) indicates the energy of initial (final) states, and $\hbar \omega$ is the photon energy.
$|\psi_g\rangle=|\psi(l_v^n)\rangle$ denotes any state of the ground configuration of the outer shell, $i.e.$ a valence electronic state of $n$ electrons with the azimuthal angular momentum $l_v$.
The final state configuration is represented by $|\psi_f\rangle=|\underline{c}_{jm}\psi^\prime(l_v^{n+1})\rangle$ where $\underline{c}_{jm}$ stands for a hole in a core level.
This formulation captures the core-level transitions induced by the electric multipole interaction, which dominates in XAS experiments.

The integral of XAS concerning the electric dipole transition from a core state with $j_\pm = l_c\pm \frac{1}{2}$ is expressed as
\begin{equation}
\begin{split} 
I^{j\pm}_\varepsilon = \int_{j_\pm} \mu_\varepsilon (\hbar\omega) d\omega= \sum_{MM^\prime}E_0^2 (-1)^{M+M^\prime}\varepsilon_{-M^\prime}^*\varepsilon_{-M}P^{j\pm}_{M^\prime M}
\end{split} 
\end{equation}
with 
\begin{equation}\label{pmm}
    P^{j\pm}_{M^\prime M} \equiv \sum_{f \in j_\pm} \langle\psi_g|r^*_{M^\prime}|\psi_f\rangle \langle\psi_f|r_{M}|\psi_g\rangle.
\end{equation}
Using the Wigner-Eckart theorem, the matrix element of photoelectron transition from the core state $jm_j$ to the valence state $l_vm_v$ by the electric dipole moment $r_{M}$ is given by
\begin{equation}\label{dipoletransition}
\langle \psi_f | r_M | \psi_g \rangle = \sum_{\{m\}} C_{l_c m_c;1M}^{l_vm_v}C_{l_c m_c;\frac{1}{2}m_s}^{jm_j}
\frac{\langle \psi_f | a_{m_vm_s}^\dagger b_{jm_{j}}|\psi_g\rangle}{\sqrt{2l_v+1}}
\langle l_v||O_1|| l_c\rangle_R
\end{equation}
where $C_{l_1m_1;l_2m_2}^{l_3m_3}$ is the Clebsh-Gordan coefficient and the index of $\{m\}$ indicates the summation for all $m$. 

Here, $b_{jm_j}$ ($b^\dagger_{jm_j}$) denotes the annihilation (creation) operator acting on core‐electron states characterized by quantum numbers $j,m_j$, whereas $a_{m_vm_s}$ ($a_{m_vm_s}^\dagger$) refers to the annihilation (creation) operator for valence states labeled by \(m_v,m_s\).
In addition, $\langle l_v||O_1|| l_c\rangle_R$ denotes the reduced matrix element, which in this paper is approximated as independent of $m$ and energy.
The Wigner–Eckart theorem establishes the selection rules for electric dipole transitions: $l_c - l_v = 0,\pm 1$, $m_c - m_v = 0,\pm 1$, and the spin of both the core and valence states is conserved.

By substituting Eq. (\ref{dipoletransition}) into Eq. (\ref{pmm}), $P^{j\pm}_{M^\prime M}$ can be rewritten as 
\begin{multline}
P^{j\pm}_{M^\prime M} = \sum_{\psi_f}
\sum_{\{m\}}
[l_v]^{-1}
C_{l_c m_c;1M}^{l_vm_v}
C_{l_c m_c;\frac{1}{2}m_s}^{jm_j}
C_{l_c m_c^\prime;1M^\prime}^{l_vm_v^\prime}
C_{l_c m_c^\prime;\frac{1}{2}m_s^\prime}^{jm_j^\prime}\\\times
\langle \psi_g | b_{jm_{j}}^\dagger a_{m_vm_s}|\psi_f\rangle
\langle \psi_f | a_{m_v^\prime m_s^\prime}^\dagger b_{jm_{j}^\prime}|\psi_g\rangle
R_{l_c}
\end{multline}
with $[l_v]\equiv 2l_v+1$ and $R_{l_c} = |\langle l_v||O_1|| l_c\rangle_R|^2$.
By taking the sum over $m_j$ while noting that $m_j^\prime = m_j$ for the absorption process, we arrive at the following equation \cite{formula1}:
\begin{equation}
P_{M^\prime M}^{j\pm}
=
\sum_{\{m\}}
\sum_{s=0}^1\sum_{n=-s}^s \frac{\alpha_\pm^{(s)}(l_c)}{[l_cl_v]}
C_{l_c m_c;1M}^{l_vm_v}
C_{l_c m_c^\prime;1M^\prime}^{l_vm_v^\prime}
C_{l_cm_c^\prime;s,-n}^{l_cm_c}
C_{\frac{1}{2}m_s^\prime;sn}^{\frac{1}{2}m_s}
R_{l_c}    
\end{equation}
with $[ab\cdots] \equiv (2a+1)(2b+1)\cdots$.
Here, $\alpha_\pm^{(s)}(l_c)$ is a coefficient depending on the core state $j_\pm = l_c\pm 1/2$ and expressed as
\begin{equation}
    \alpha^{(s)}_\pm(l_c) = \begin{cases}
    j_\pm+\frac{1}{2} & (s=0) \\
    \pm \sqrt{3l_c(l_c+1)} & (s=1).  
  \end{cases}
\end{equation}
This parameter implies that the absorption intensity can be divided into processes in which the spin does not flip $(s=0)$ and those where the spin flips $(s=1)$.
By summing the absorption intensities from both $j_\pm$, the spin-flip processes cancel out, leaving only the information from the spin nonflip process.
By calculating the absorption intensity for $j_\pm$ separately and taking the appropriate difference, the information on the spin-flop process can be obtained.
In other words, observing the absorption process of the spin-flop transition requires absorption in the inner core levels split by the spin-orbit interaction.

\section{Polarization Sum rules of X-ray absorption}
To relate X-ray absorption to the physical symmetry of the holes in the valence state, here, we introduce a multipole with the quantum numbers $l$ and $m$ defined as
\begin{equation}\label{slm}
    S_{lm}^{j\pm}  \equiv \sum_{MM^\prime} C_{1M^\prime;lm}^{1M}P_{M^\prime M}^{j\pm},
\end{equation}
giving some descriptions of the sum of XAS with different X-ray polarizations.
The sum of XAS for isotropic polarization is linked to the electric monopole, which is confirmed by $S_{00}^{j\pm} = I_z^{j\pm}+ I_+^{j\pm}+I_-^{j\pm}$ due to $C_{1M^\prime;00}^{1M}=\delta_{MM^\prime}$.
The dipole moment quantum number gives $S_{10}^{j\pm} = \frac{1}{\sqrt{2}}(I_-^{j\pm}- I_+^{j\pm})$ due to $C_{1M^\prime;10}^{1M}=\frac{1}{\sqrt{2}}M\delta_{MM^\prime}$, which corresponds to XMCD when the incident X-ray is aligned along the $z$ axis.
In addition, we consider (i) perpendicular and (ii) in-plane X-ray magnetic linear dichroism (XMLD).
The perpendicular XMLD is expressed by
a quadrupole moment $S_{20}^{j\pm} = \frac{1}{\sqrt{10}}(I_+^{j\pm} + I_-^{j\pm} - 2I_z^{j\pm})$,
representing the difference in XAS with linear polarization between the perpendicular and in-plane directions.
In contrast, the in-plane XMLD, i.e., the difference in XAS with linear polarization between the in-plane $x$- and $y$-directions, can not be described by a single $S_{lm}$.
However, using a relation of $S_{2,\pm 2}^{j\pm} = \frac{3}{\sqrt{15}}P_{\pm 1, \mp 1}^{j\pm}$,
it is confirmed that its difference is linked to the in-plane XMLD as
$S_{2,2}^{j\pm} + S_{2,-2}^{j\pm} = \frac{3}{\sqrt{15}}(I_y^{j\pm} - I_x^{j\pm})$ as discussed below.

Substituting Eq. (\ref{pmm}) into Eq. (\ref{slm}) results in a product of four Clebsch-Gordan coefficients involving summation over $m_c, m_c^\prime, M$, and $M^\prime$.
By applying the transformation formula for the Wigner-9$j$ symbol \cite{formulra3}, Equation (\ref{slm}) can be rewritten as

\begin{equation}
S_{lm}^{j\pm}=
\sum_{s,\kappa}
\sum_{n,\mu}
\sum_{\{m\}}
\alpha^{(s)}_\pm\beta^{(s,\kappa)}_{l}
\langle \psi_g|
C_{\kappa\mu;sn}^{lm}
C_{l_vm_v^\prime;\kappa\mu}^{l_vm_v}
C_{\frac{1}{2}m_s^\prime;sn}^{\frac{1}{2}m_s}
a_{m_vm_s} a_{m_v^\prime m_s^\prime}^\dagger|\psi_g\rangle
\end{equation}
where $s=0,1$ and $|s-l|\le \kappa \le \min (2l_v, s+l)$ ($\kappa$ is integer).
The coefficient $\beta_{l}^{(s,\kappa)}$ is independent of any $m$ and expressed using the Wigner-$9j$ symbol as
\begin{equation}\label{wig9}
\beta^{(s,\kappa)}_{l}(l_c,l_v,L)\equiv \frac{[\kappa] 
[L]^{\frac{1}{2}}}{[l_cl_vl]^{\frac{1}{2}}}
\begin{Bmatrix}
l_v & l_c & L \\
l_v & l_c & L \\
\kappa & s & l
\end{Bmatrix} R_{l_c}
\end{equation}
where $[ab\cdots]^\frac{1}{2} \equiv \{(2a+1)(2b+1)\cdots\}^\frac{1}{2}$ and $\{\ddots\}$ indicate the Wigner-9$j$ symbols.
Here, we consider a complete multipole operator for holes in the valence of the ground state $|\psi_g\rangle$ expressed by $\underline{\hat{X}}_{lm}^{(s,k)}$ based on the analogy of Eq. (\ref{multipole}).
Consequently, the sum of XAS at $j_{c\pm} = l_c\pm \frac{1}{2}$ core state can be rewritten by using the expectation value of complete magnetic multipoles for holes and Eq. (\ref{kappamu}) as
\begin{equation}
    S_{lm}^{j\pm}
    =\alpha_{\pm}^{(0)}\beta_{l}^{(0,l)}   \langle \underline{X}_{lm}^{(0,0)} \rangle 
    +\alpha_{\pm}^{(1)}\beta_{l}^{(1,l-1)} \langle \underline{X}_{lm}^{(1,-1)}\rangle
    -\alpha_{\pm}^{(1)}\beta_{l}^{(1,l+1)} \langle \underline{X}_{lm}^{(1,1)} \rangle,
\end{equation}
with $$\langle \underline{X}_{lm}^{(s,k)}\rangle \equiv 
\sum_{n,\mu}
\sum_{\{m\}}i^{s+k}
\langle \psi_g|
C_{\kappa\mu;sn}^{lm}
C_{l_vm_v^\prime;\kappa\mu}^{l_vm_v}
C_{\frac{1}{2}m_s^\prime;sn}^{\frac{1}{2}m_s}
a_{m_vm_s} a_{m_v^\prime m_s^\prime}^\dagger|\psi_g\rangle.$$
Since $\langle \underline{X}_{lm}^{(s,k)}\rangle$ contains only the expectation value for angular information of the complete multipole basis, it can be represented by equivalent operators \cite{carra1993magnetic}.
Noted that the Wigner-9$j$ symbol in Eq. (\ref{wig9}) becomes zero when $s = 1$ and $k =l$, and therefore $\beta_{l}^{(1,l)}=0$.
This means that the sum rule of dipole transition can capture the information for electric and magnetic multipoles, $\bm{Q}$ and $\bm{M}$ on a complete magnetic multipole basis, and cannot directly apply to electric and magnetic toroidal moments, $\bm{G}$ and $\bm{T}$.

Using the relations $\alpha_+^{(1)}+\alpha_-^{(1)}=0$, the sum over $S_{lm}^{j\pm}$ on the two core states $j_{\pm} = l_c\pm \frac{1}{2}$ is expressed as,
\begin{equation}
    S_{lm}^{j+} + S_{lm}^{j-}= [l_c]\beta_{l}^{(0,l)} \langle \underline{X}_{lm}^{(0,0)}\rangle,
\end{equation}
suggesting that it gives expectation values for the spinless multipoles.
Hereafter, it will be referred to as the spinless sum rule.
Similarly, by using a relation $\alpha_+^{(0)}-\frac{l_c+1}{l_c}\alpha_-^{(0)}=0$, another sum rule can be derived, 
\begin{equation}
    S_{lm}^{j+} -\frac{l_c+1}{l_c} S_{lm}^{j-}= [l_c]\sqrt{\frac{3(l_c+1)}{l_c}}
     \left\{\beta_{l}^{(1,l-1)} \langle \underline{X}_{lm}^{(1,-1)}\rangle
    -\beta_{l}^{(1,l+1)} \langle \underline{X}_{lm}^{(1,1)} \rangle\right\},
\end{equation}
which allows for the detection of only spinful multipoles.
Hereafter, it will be referred to as the spinful sum rule.
In other words, depending on the method of calculating the absorption sum with the core as the reference, it is possible to extract either spinless or spinful magnetic multipoles selectively.

\section{Explicit expression of magnetic multipoles in sum rules}
In the present paper, we will examine the relationship between the XAS, XMCD, and XMLD sum rules, and magnetic multipoles concretely using the case of the dipole transition $l_c=1$ ($2p$ orbitals) $\rightarrow$ $l_v=2$ ($3d$ orbitals) as an example.
The relation between the polarization sum rules and the complete magnetic multipole basis and physical quantity is classified in Table \ref{technique_property}.

\subsection{Monopole sum rule}
The sum rule on the isotropic XAS ($I^{j\pm}_{\rm XAS}=I_z^{j\pm}+I_+^{j\pm}+I_-^{j\pm}$) gives the complete multipole basis of charge monopole \( l = 0, m = 0 \). 
Since the spinless monopole is given by $\underline{Q}_{00}^{(0,0)} = \underline{Q}_{00}^{\text{(orb)}}=O_{00}$, the spinless sum rule ($s=0$) is expressed as 
\begin{equation}
    I_{\rm XAS}^{j+} + I_{\rm XAS}^{j-}= \langle n_h \rangle C,
\end{equation}
where $n_h$ indicates the number of holes in the $l_v$ state and $C$ is the normalized constant factor including the radial matrix element of the dipole transition and multipoles.
On the other hand, the spinful monopole is given by 
\begin{equation}
{Q}_{00}^{(1,1)}=-\sum_{n}C_{1,-n;1,n}^{00}{M}^{\text{(orb)}}_{1,-n}\sigma_{1n}    
 = \frac{1}{\sqrt{3}}(\hat{\bm{l}}\cdot \hat{\bm{s}}),
\end{equation}
which involves the spin-orbit coupling  ($\lambda \hat{\bm{l}}\cdot \hat{\bm{s}}$) in the $l_v$ state.
Consequently, the intensity of the spinful monopole sum rule is proportional to ${Q}_{00}^{(1,1)}$, and is expressed as
\begin{equation}
    I_{\rm XAS}^{j+} -2 I_{\rm XAS}^{j-}= \langle \hat{\underline{\bm{l}}}\cdot \underline{\hat{\bm{s}}}\rangle C, 
\end{equation}
where $C$ is the same normalized factor as in Eq. (23).
This means that when the spin–orbit interaction in $l_v$ is zero, the intensity ratio of $I^{j+}_{\rm XAS}$ to $I^{j-}_{\rm XAS}$ is 2:1, and conversely, as the interaction becomes stronger, the absorption intensity of $I^{j-}_{\rm XAS}$ decreases.

These results suggest that spinful electric dipole transitions offer a pathway to probe the spin-orbit interaction in the valence states. 
However, in early $3d$ transition metals, such as V and Cr, the $L_3$/$L_2$ branching ratio significantly deviates from the expected 2:1 value, tending toward 1:1 due to strong electron-core-hole interaction and the small spin-orbit splitting of the $2p$ core levels \cite{Ankudinov2003-vd, Schwitalla1998-ro}. 
This highlights the need for caution when interpreting branching ratios in such systems.

\subsection{Dipole sum rule}
The sum rule for $l=1$ represents that for the XMCD spectrum  ($I^{j\pm}_{\rm XMCD}=I_+^{j\pm}-I_-^{j\pm}$) and involves the first-order multipoles associated with the magnetic dipole moment.
Here, we consider $m =0$, that is, $S_{10}^{j\pm} = -\frac{1}{\sqrt{2}}(I_+^{j\pm}- I_-^{j\pm})$, which gives the component of each dipole moment projected onto the $z$ direction.
Since the spinless dipole moment is given by $\underline{M}_{10}^{(0,0)} = \underline{M}_{10}^{\text{(orb)}} = \underline{l}_z$, the spinless sum rule ($s=0$) is expressed as 
\begin{equation}
    I^{j+}_{\rm XMCD} + I^{j-}_{\rm XMCD} =  -\frac{1}{2} \langle \underline{l}_z\rangle C,
\end{equation}
indicating that the sum rule of the XMCD selectively extracts information on orbital angular momentum in magnetization \cite{Thole1992}. 

On the other hand, spinful dipole moments have two components: one is the pure spin represented by $\underline{M}_{10}^{(1,-1)} = C^{10}_{00;10}\underline{Q}_{00}^{\text{(orb)}}\underline{\sigma}_z = \underline{\sigma}_z$, and the other is a term arising from the coupling between the electric quadrupole and the spin, 
\begin{equation}
    [M_{10}^{(1,1)}]_z=-\sum_{n}C^{10}_{2,-n;1n}Q^{\text{(orb)}}_{2,-n}\sigma_{1n} = \frac{1}{\sqrt{10}}\sum_i (3z r_i - \bm{r}^2\delta_{zr_i})\sigma_z
\end{equation}
with $\bm{r}$ being operators for the unit vector of position \cite{hayami2024unified}.
The latter is proportional to the anisotropic magnetic dipole $t_z$ term \cite{Oguchi}, which provides information on the anisotropy of the electron spin-density distribution.
Consequently, the spinful sum rule ($s=1$) is expressed as
\begin{equation}
    I^{j+}_{\text{XMCD}} -2 I^{j-}_{\text{XMCD}}= -\left(\frac{2}{3}\langle \underline{s}_z\rangle + \frac{7}{3}\langle \underline{t}_z\rangle\right)C, 
\end{equation}
with $t_z = \frac{1}{4}\left(3[l_z(\bm{l}\cdot\bm{s})]_+-2l^2s_z\right)$ in the equivalent $l$ orperator form \cite{PhysRevB.57.112}.

XMCD is used to separate spin and orbital contributions, and is particularly valuable in ferrimagnetic materials because it allows element-specific evaluation \cite{Bitla2015-hs}.
Additionally, the spinful multipole term $t_z$ has been used to uncover the role of hidden multipoles in magnetic materials. 
For example, in an exchange‐bias Fe/MgO system, an electric‐field–driven XMCD response has been observed, which originates from changes in the anisotropic spatial distribution of spin involved with the anisotropic magnetic dipole term $t_z$ \cite{Miwa2015APL}.
The $t_z$ component of the spinful dipole is a good descriptor for the characteristics of the exchange bias effect and its response to an electric field.
Additionally, the $t_z$ term has recently been identified as a key origin of XMCD signals in antiferromagnets lacking net magnetization, as exemplified by chiral antiferromagnetic Mn$_3$Sn \cite{Yamasaki2020, Sasabe_PRL_Mn3Sn} and collinear antiferromagnetic (altermagnetic) systems, such as RuO$_2$ \cite{sasabe2023ferroic} and MnTe \cite{amin2024nanoscale}.
Magnetic materials that exhibit XMCD despite being antiferromagnetic can be classified, within a magnetic multipole basis, as spinful magnetic dipoles; adopting this framework is essential both for categorizing such materials and for interpreting their behavior. 
Indeed, it has even been theoretically proposed 
that the spinful magnetic dipole $t_z$ underlies the emergence of anomalous Hall effect in antiferromagnet \cite{hayami2021essential}.

\subsection{Quadrupole sum rule for out-of-plane XMLD (zXMLD)}
The sum rule for $l=2$ represents that for the XMLD spectrum and involves the second-order multipoles associated with the electronic quadrupole moment.
First, we consider $m = 0$, that is, $S_{20}^{j\pm} = \frac{1}{\sqrt{10}}(I_+^{j\pm} + I_-^{j\pm} - 2I_z^{j\pm})$, which shows the difference in x-ray absorption intensity when the polarization is applied within the $xy$ plane and along the out-of-plane direction [see Fig. 1(a) and (b)].
The spinless multipole of the quadrupole moment is given by $\underline{Q}_{20}^{(0,0)} = \underline{Q}_{20}^{\text{(orb)}}$, which represents the electronic quadrupole moment $O_{20}=\frac{1}{2}(3z^2-r^2)$, corresponding to $d_{3z^2-r^2}$ orbital.
Therefore, the spinless sum rule ($s=0$) is expressed as,
\begin{equation}
     I_{\text{zXMLD}}^{j+} + I_{\text{zXMLD}}^{j-}= \langle \underline{Q}_{zz}\rangle C,
\end{equation}
with the equivalent operator $Q_{zz} = \frac{1}{2}(l_z^2-\frac{1}{3}l^2)$ \cite{carra1993magnetic}.
This result reflects the anisotropy of charge distribution between the out-of-plane and the in-plane direction.

The spinful electronic quadrupole moments are composed of magnetic dipole and octupole terms.
The magnetic dipole one is given by 
\begin{equation}
 Q_{20}^{(1,-1)}=\sum_{n}C_{1,-n;1n}^{2,0}M^{\text{(orb)}}_{1,-n}\sigma_{1n}=\frac{1}{\sqrt{6}} (3l_zs_z-\bm{l}\cdot\bm{s}),
\end{equation}
indicating the anisotropy of the spin-orbit coupling ($\lambda \bm{l}\cdot\bm{s}$), which enables the probing of the magnetocrystalline anisotropy \cite{van1999magnetic}.
The equivalent operator is given by $P_{zz} = \frac{1}{2}(3l_zs_z-\bm{l\cdot s})$ \cite{carra1993magnetic,PhysRevB.57.112}.
The octupole term in the spinful quadrupole moment is expressed as
\begin{equation}
 Q_{20}^{(1,1)}=-\sum_{n}C_{3,-n;1n}^{2,0}M^{\text{(orb)}}_{3,-n}\sigma_{1n}
\end{equation}
where the magnetic octupole operators are shown in the Table \ref{nabla_O_lm}. 
For example, when the spin is oriented along the $z$-axis, the explicit expression by extracting only the term proportional to $s_z$ is given by 
\begin{equation}
 M_{30}^{\text{(orb)}}\sigma_z =-\frac{3}{2}[zxl_x+zyl_y]_+ s_z + \frac{3}{2} (3z^2 - r^2) l_zs_z
\end{equation}
which reflects a quantity combining the electric quadrupole and spin-orbit coupling.
Consequently, the spinful sum rule is expressed as,
\begin{equation}
I^{j+}_{\text{zXMLD}} -2 I^{j-}_{\text{zXMLD}}= \left(\frac{2}{5}\langle \underline{P_{zz}}\rangle + \frac{3}{5}\langle \underline{R_{zz}}\rangle\right)C,
\end{equation}
where $R_{zz} = \frac{1}{3}[5l_z(\bm{l\cdot s})l_z-(l^2-2)\bm{l\cdot s} -(2l^2+1)l_zs_z]$ is the equivalent operator of $Q_{20}^{(1,1)}$ \cite{carra1993magnetic,PhysRevB.57.112}.
In the case of the electric quadrupole sum rule, it is difficult to separately measure the spinful magnetic dipole and magnetic octupole. 
As a result, it is also useful to express as $I^{j+}_{\text{zXMLD}} -2 I^{j-}_{\text{zXMLD}}=\langle \underline{U}_{zz}\rangle C$ using the total operator $U_{zz}= l_z(\bm{l\cdot s})l_z-2l_zs_z-\bm{l\cdot s}$.

Out-of-plane XMLD has been used to determine magnetic anisotropy, for example in perpendicularly magnetized Fe/MgO films \cite{Okabayashi2014-rp} and Mn$_{3-\delta}$Ga alloys \cite{Okabayashi2020-dn}. 
Since it can detect magnetic multipoles corresponding to the anisotropic components of the spin–orbit interaction \cite{van1999magnetic}, it holds promise for estimating the perpendicular magnetic anisotropy energy based on the spinful electronic quadrupole multipoles. 
Moreover, when a magnetic field is applied to a magnetic octupole state, the response corresponds to $R_{zz}$ multipole, suggesting potential applicability for detection and quantitative evaluation of magnetic octupole order through the XMLD sum rule such as in CeB$_6$ \cite{Matsumura2009-kg}.

\subsection{Quadrupole sum rule for in-plane XMLD (xyXMLD)}
Next, we consider $l=2$ and $m=\pm 2$ cases, $S_{2,\pm 2}^{j\pm} = \frac{3}{\sqrt{15}}P_{\pm 1, \mp 1}^{j\pm}$, corresponding to the cross term of left and right circular polarization, which is not possible to directly observe in absorption which can only describe the polarization of the incident x-rays as shown in Equation (\ref{fermi}). 
Therefore, we consider the combination of sum rules for $m=\pm 2$, and then it becomes clear that the signal can be observed as an in-plane xyXMLD [$I_{\text{xyXMLD}}^{j\pm} = I_x^{j\pm}-I_y^{j\pm}$] component using $S_{2,2}^{j\pm} + S_{2,-2}^{j\pm} = \frac{3}{\sqrt{15}}(I_x^{j\pm} - I_y^{j\pm})$.
The corresponding spinless multipole of quadrupole moment is given by $\underline{Q}_{2,2}^{\text{(orb)}}+\underline{Q}_{2,-2}^{\text{(orb)}}$; consequently, the spinless sum rule is expressed as
\begin{equation}
     I_{\text{xyXMLD}}^{j+} + I_{\text{xyXMLD}}^{j-}= \langle \underline{Q}_{x^2-y^2} \rangle C,
\end{equation}
with the operator equivalent $Q_{x^2-y^2}\equiv (l_y^2-l_x^2)/6$, reflecting the anisotropic charge distribution within the $xy$-plane.

The spinful electronic quadrupole moments are composed of magnetic dipole and octupole terms.
The magnetic dipole one is given by 
\begin{equation}
    Q_{x^2-y^2}^{(1,-1)}\equiv Q_{2,2}^{(1,-1)}+Q_{2,-2}^{(1,-1)} = \frac{1}{2} (l_xs_x-l_ys_y)
\end{equation}
with $Q_{2,\pm 2}^{(1,-1)}=\sum_n C_{1,\pm 2-n;1n}^{2,\pm 2}M^{\text{(orb)}}_{1,\pm 2-n}\sigma_{1n}$, indicating the anisotropy of the spin-orbit coupling within the $xy$-plane.
On the other hand, the octupole term is expressed as
\begin{equation}
 Q_{2,\pm 2}^{(1,1)}=-\sum_{n}C_{3,\pm 2 -n;1n}^{2,\pm 2}M^{\text{(orb)}}_{3,\pm 2-n}\sigma_{1n}
\end{equation}
where the magnetic octupole operators are given in Table \ref{nabla_O_lm}.
For example, when the spin is oriented along the $z$-axis, extracting only the term proportional to $s_z$ results in 
\begin{equation}
(M^{\text{(orb)}}_{3,2}+M^{\text{(orb)}}_{3,-2})\sigma_{z} = \sqrt{\frac{15}{2}} [xz l_x- yzl_y]_+s_z+(x^2- y^2)l_zs_z,
\end{equation}
which reflects a quantity combining the electric quadrupole and spin-orbit coupling involving the in-plane anisotropy.
Consequently, the spinful sum rule for xyXMLD is expressed as
\begin{equation}
     I_{\text{xyXMLD}}^{j+} -2 I_{\text{xyXMLD}}^{j-}= \left(\frac{2}{5}\langle \underline{P}_{x^2-y^2}\rangle + \frac{3}{5}\langle \underline{R}_{x^2-y^2}\rangle\right)C,
\end{equation}
with the operator equivalents $P_{x^2-y^2} \equiv \frac{2}{3}(l_xs_x-l_ys_y)$ and $R_{x^2-y^2} \equiv \frac{2}{9}\{l_y(\bm{l}\cdot\bm{s})l_y - l_x(\bm{l}\cdot\bm{s})l_x\}$.
These operators reflect the in-plane anisotropy of the spin–orbit interaction.

For example, such in-plane XMLD has been employed to visualize the spatial domain structure of the N\'{e}el vector in antiferromagnetic NiO \cite{Toyohiko2012-ck}. 
With the present formalism, it is expected that information on anisotropic spin–orbit interactions can be extracted from the integrated XMLD spectrum, providing crucial insight for evaluating the magnetic anisotropy energy of antiferromagnetic materials. 
Additionally, since these electric quadrupole moments have the same symmetry as the magnetic toroidal quadrupole moment with the applied magnetic field, it is expected to detect the anisotropy of the electronic states in $d$-wave altermagnets such as in MnF$_2$ \cite{higuchi2016control} and NiCo$_2$O$_4$ \cite{koizumi2023quadrupole}.

\section{Conclusion}
In this study, we have reconsidered the sum rules for X-ray absorption spectroscopy based on a complete magnetic multipole basis. 
We have demonstrated that it naturally derives the anisotropic magnetic dipole term in the XMCD, which plays important role in the $s$-wave altermagnetic system.
Additionally, we have shown that the sum rules for out-of-plane and in-plane X-ray Magnetic Linear Dichroism (XMLD) can be derived using the electric quadrupole contributions. 
This approach provides a unified theoretical framework to analyze and interpret dichroic signals in a wide range of magnetic materials.
Our findings not only enhance the understanding of XMCD and XMLD but also pave the way for further research in complex magnetic structures. 
Future studies incorporating this multipole-based methodology could further refine our insights into spin-orbit interactions and hidden magnetic orderings in advanced magnetic materials.

\section{Acknowledgment}
The authors thank T. Arima and M. Mizumaki for the productive discussion.
This project is partly supported by the Japan Society for the Promotion of Science (JSPS) KAKENHI (19H04399, 24K03205, and 24H01685).
This work was supported by MEXT Quantum Leap Flagship Program (MEXT Q-LEAP) Grant Number JPMXS0118068681.
This work is also partially supported by CREST(JPMJCR1861 and JPMJCR2435), Japan Science and Technology Agency (JST).

\bibliographystyle{apsrev4-2}

\begin{thebibliography}{53}%
\makeatletter
\providecommand \@ifxundefined [1]{%
 \@ifx{#1\undefined}
}%
\providecommand \@ifnum [1]{%
 \ifnum #1\expandafter \@firstoftwo
 \else \expandafter \@secondoftwo
 \fi
}%
\providecommand \@ifx [1]{%
 \ifx #1\expandafter \@firstoftwo
 \else \expandafter \@secondoftwo
 \fi
}%
\providecommand \natexlab [1]{#1}%
\providecommand \enquote  [1]{``#1''}%
\providecommand \bibnamefont  [1]{#1}%
\providecommand \bibfnamefont [1]{#1}%
\providecommand \citenamefont [1]{#1}%
\providecommand \href@noop [0]{\@secondoftwo}%
\providecommand \href [0]{\begingroup \@sanitize@url \@href}%
\providecommand \@href[1]{\@@startlink{#1}\@@href}%
\providecommand \@@href[1]{\endgroup#1\@@endlink}%
\providecommand \@sanitize@url [0]{\catcode `\\12\catcode `\$12\catcode `\&12\catcode `\#12\catcode `\^12\catcode `\_12\catcode `\%12\relax}%
\providecommand \@@startlink[1]{}%
\providecommand \@@endlink[0]{}%
\providecommand \url  [0]{\begingroup\@sanitize@url \@url }%
\providecommand \@url [1]{\endgroup\@href {#1}{\urlprefix }}%
\providecommand \urlprefix  [0]{URL }%
\providecommand \Eprint [0]{\href }%
\providecommand \doibase [0]{https://doi.org/}%
\providecommand \selectlanguage [0]{\@gobble}%
\providecommand \bibinfo  [0]{\@secondoftwo}%
\providecommand \bibfield  [0]{\@secondoftwo}%
\providecommand \translation [1]{[#1]}%
\providecommand \BibitemOpen [0]{}%
\providecommand \bibitemStop [0]{}%
\providecommand \bibitemNoStop [0]{.\EOS\space}%
\providecommand \EOS [0]{\spacefactor3000\relax}%
\providecommand \BibitemShut  [1]{\csname bibitem#1\endcsname}%
\let\auto@bib@innerbib\@empty
\bibitem [{\citenamefont {de~Groot}\ and\ \citenamefont {Kotani}(2008)}]{Groot2008}%
  \BibitemOpen
  \bibfield  {author} {\bibinfo {author} {\bibfnamefont {F.~M.~F.}\ \bibnamefont {de~Groot}}\ and\ \bibinfo {author} {\bibfnamefont {A.}~\bibnamefont {Kotani}},\ }\href@noop {} {\emph {\bibinfo {title} {Core Level Spectroscopy of Solids}}}\ (\bibinfo  {publisher} {Cambridge University Press},\ \bibinfo {year} {2008})\ \bibinfo {note} {iSBN: 978-0521831793}\BibitemShut {NoStop}%
\bibitem [{\citenamefont {Sch{\"u}tz}\ \emph {et~al.}(1987)\citenamefont {Sch{\"u}tz}, \citenamefont {Wagner}, \citenamefont {Wilhelm}, \citenamefont {Kienle}, \citenamefont {Zeller}, \citenamefont {Frahm},\ and\ \citenamefont {Materlik}}]{Schuetz1987}%
  \BibitemOpen
  \bibfield  {author} {\bibinfo {author} {\bibfnamefont {G.}~\bibnamefont {Sch{\"u}tz}}, \bibinfo {author} {\bibfnamefont {W.}~\bibnamefont {Wagner}}, \bibinfo {author} {\bibfnamefont {W.}~\bibnamefont {Wilhelm}}, \bibinfo {author} {\bibfnamefont {P.}~\bibnamefont {Kienle}}, \bibinfo {author} {\bibfnamefont {R.}~\bibnamefont {Zeller}}, \bibinfo {author} {\bibfnamefont {R.}~\bibnamefont {Frahm}},\ and\ \bibinfo {author} {\bibfnamefont {G.}~\bibnamefont {Materlik}},\ }\href@noop {} {\bibfield  {journal} {\bibinfo  {journal} {Phys. Rev. Lett.}\ }\textbf {\bibinfo {volume} {58}},\ \bibinfo {pages} {737} (\bibinfo {year} {1987})}\BibitemShut {NoStop}%
\bibitem [{\citenamefont {St{\"o}hr}(1999)}]{Stohr1999}%
  \BibitemOpen
  \bibfield  {author} {\bibinfo {author} {\bibfnamefont {J.}~\bibnamefont {St{\"o}hr}},\ }\href@noop {} {\bibfield  {journal} {\bibinfo  {journal} {J. Magn. Magn. Mater.}\ }\textbf {\bibinfo {volume} {200}},\ \bibinfo {pages} {470} (\bibinfo {year} {1999})}\BibitemShut {NoStop}%
\bibitem [{\citenamefont {Alders}\ \emph {et~al.}(1998)\citenamefont {Alders}, \citenamefont {Tjeng}, \citenamefont {Voogt}, \citenamefont {Hibma}, \citenamefont {Sawatzky}, \citenamefont {Chen}, \citenamefont {Vogel}, \citenamefont {Sacchi},\ and\ \citenamefont {Iacobucci}}]{Alders1998}%
  \BibitemOpen
  \bibfield  {author} {\bibinfo {author} {\bibfnamefont {D.}~\bibnamefont {Alders}}, \bibinfo {author} {\bibfnamefont {L.~H.}\ \bibnamefont {Tjeng}}, \bibinfo {author} {\bibfnamefont {F.~C.}\ \bibnamefont {Voogt}}, \bibinfo {author} {\bibfnamefont {T.}~\bibnamefont {Hibma}}, \bibinfo {author} {\bibfnamefont {G.~A.}\ \bibnamefont {Sawatzky}}, \bibinfo {author} {\bibfnamefont {C.~T.}\ \bibnamefont {Chen}}, \bibinfo {author} {\bibfnamefont {J.}~\bibnamefont {Vogel}}, \bibinfo {author} {\bibfnamefont {M.}~\bibnamefont {Sacchi}},\ and\ \bibinfo {author} {\bibfnamefont {S.}~\bibnamefont {Iacobucci}},\ }\href@noop {} {\bibfield  {journal} {\bibinfo  {journal} {Phys. Rev. B}\ }\textbf {\bibinfo {volume} {57}},\ \bibinfo {pages} {11623} (\bibinfo {year} {1998})}\BibitemShut {NoStop}%
\bibitem [{\citenamefont {van~der Laan}\ and\ \citenamefont {Figueroa}(2014)}]{vanDerLaan2014}%
  \BibitemOpen
  \bibfield  {author} {\bibinfo {author} {\bibfnamefont {G.}~\bibnamefont {van~der Laan}}\ and\ \bibinfo {author} {\bibfnamefont {A.~I.}\ \bibnamefont {Figueroa}},\ }\href@noop {} {\bibfield  {journal} {\bibinfo  {journal} {Coord. Chem. Rev.}\ }\textbf {\bibinfo {volume} {277--278}},\ \bibinfo {pages} {95} (\bibinfo {year} {2014})}\BibitemShut {NoStop}%
\bibitem [{\citenamefont {Thole}\ \emph {et~al.}(1992)\citenamefont {Thole}, \citenamefont {Carra}, \citenamefont {Sette},\ and\ \citenamefont {van~der Laan}}]{Thole1992}%
  \BibitemOpen
  \bibfield  {author} {\bibinfo {author} {\bibfnamefont {B.~T.}\ \bibnamefont {Thole}}, \bibinfo {author} {\bibfnamefont {P.}~\bibnamefont {Carra}}, \bibinfo {author} {\bibfnamefont {F.}~\bibnamefont {Sette}},\ and\ \bibinfo {author} {\bibfnamefont {G.}~\bibnamefont {van~der Laan}},\ }\href@noop {} {\bibfield  {journal} {\bibinfo  {journal} {Phys. Rev. Lett.}\ }\textbf {\bibinfo {volume} {68}},\ \bibinfo {pages} {1943} (\bibinfo {year} {1992})}\BibitemShut {NoStop}%
\bibitem [{\citenamefont {Carra}\ \emph {et~al.}(1993{\natexlab{a}})\citenamefont {Carra}, \citenamefont {Thole}, \citenamefont {Altarelli},\ and\ \citenamefont {Wang}}]{Carra1993}%
  \BibitemOpen
  \bibfield  {author} {\bibinfo {author} {\bibfnamefont {P.}~\bibnamefont {Carra}}, \bibinfo {author} {\bibfnamefont {B.~T.}\ \bibnamefont {Thole}}, \bibinfo {author} {\bibfnamefont {M.}~\bibnamefont {Altarelli}},\ and\ \bibinfo {author} {\bibfnamefont {X.}~\bibnamefont {Wang}},\ }\href@noop {} {\bibfield  {journal} {\bibinfo  {journal} {Phys. Rev. Lett.}\ }\textbf {\bibinfo {volume} {70}},\ \bibinfo {pages} {694} (\bibinfo {year} {1993}{\natexlab{a}})}\BibitemShut {NoStop}%
\bibitem [{\citenamefont {Chen}\ \emph {et~al.}(1995)\citenamefont {Chen}, \citenamefont {Idzerda}, \citenamefont {Lin}, \citenamefont {Smith}, \citenamefont {Meigs}, \citenamefont {Chaban}, \citenamefont {Ho}, \citenamefont {Pellegrin},\ and\ \citenamefont {Sette}}]{Chen1995}%
  \BibitemOpen
  \bibfield  {author} {\bibinfo {author} {\bibfnamefont {C.~T.}\ \bibnamefont {Chen}}, \bibinfo {author} {\bibfnamefont {Y.~U.}\ \bibnamefont {Idzerda}}, \bibinfo {author} {\bibfnamefont {H.-J.}\ \bibnamefont {Lin}}, \bibinfo {author} {\bibfnamefont {N.~V.}\ \bibnamefont {Smith}}, \bibinfo {author} {\bibfnamefont {G.}~\bibnamefont {Meigs}}, \bibinfo {author} {\bibfnamefont {E.}~\bibnamefont {Chaban}}, \bibinfo {author} {\bibfnamefont {G.~H.}\ \bibnamefont {Ho}}, \bibinfo {author} {\bibfnamefont {E.}~\bibnamefont {Pellegrin}},\ and\ \bibinfo {author} {\bibfnamefont {F.}~\bibnamefont {Sette}},\ }\href@noop {} {\bibfield  {journal} {\bibinfo  {journal} {Phys. Rev. Lett.}\ }\textbf {\bibinfo {volume} {75}},\ \bibinfo {pages} {152} (\bibinfo {year} {1995})}\BibitemShut {NoStop}%
\bibitem [{\citenamefont {Teramura}\ \emph {et~al.}(1996)\citenamefont {Teramura}, \citenamefont {Tanaka},\ and\ \citenamefont {Jo}}]{TanakaJo1996JPSJ}%
  \BibitemOpen
  \bibfield  {author} {\bibinfo {author} {\bibfnamefont {Y.}~\bibnamefont {Teramura}}, \bibinfo {author} {\bibfnamefont {A.}~\bibnamefont {Tanaka}},\ and\ \bibinfo {author} {\bibfnamefont {T.}~\bibnamefont {Jo}},\ }\href@noop {} {\bibfield  {journal} {\bibinfo  {journal} {Journal of the Physical Society of Japan}\ }\textbf {\bibinfo {volume} {65}},\ \bibinfo {pages} {1053} (\bibinfo {year} {1996})}\BibitemShut {NoStop}%
\bibitem [{\citenamefont {Bi}\ \emph {et~al.}(2014)\citenamefont {Bi}, \citenamefont {Liu}, \citenamefont {Newhouse-Illige}, \citenamefont {Xu}, \citenamefont {Rosales}, \citenamefont {Freeland}, \citenamefont {Mryasov}, \citenamefont {Zhang}, \citenamefont {te~Velthuis},\ and\ \citenamefont {Wang}}]{Chong2014PRL}%
  \BibitemOpen
  \bibfield  {author} {\bibinfo {author} {\bibfnamefont {C.}~\bibnamefont {Bi}}, \bibinfo {author} {\bibfnamefont {Y.}~\bibnamefont {Liu}}, \bibinfo {author} {\bibfnamefont {T.}~\bibnamefont {Newhouse-Illige}}, \bibinfo {author} {\bibfnamefont {M.}~\bibnamefont {Xu}}, \bibinfo {author} {\bibfnamefont {M.}~\bibnamefont {Rosales}}, \bibinfo {author} {\bibfnamefont {J.~W.}\ \bibnamefont {Freeland}}, \bibinfo {author} {\bibfnamefont {O.}~\bibnamefont {Mryasov}}, \bibinfo {author} {\bibfnamefont {S.}~\bibnamefont {Zhang}}, \bibinfo {author} {\bibfnamefont {S.~G.~E.}\ \bibnamefont {te~Velthuis}},\ and\ \bibinfo {author} {\bibfnamefont {W.~G.}\ \bibnamefont {Wang}},\ }\href@noop {} {\bibfield  {journal} {\bibinfo  {journal} {Phys. Rev. Lett.}\ }\textbf {\bibinfo {volume} {113}},\ \bibinfo {pages} {267202} (\bibinfo {year} {2014})}\BibitemShut {NoStop}%
\bibitem [{\citenamefont {Zhang}\ \emph {et~al.}(2015)\citenamefont {Zhang}, \citenamefont {Sun}, \citenamefont {L\"u}, \citenamefont {Venkatesan}, \citenamefont {Han}, \citenamefont {Zhu}, \citenamefont {Chen},\ and\ \citenamefont {Chow}}]{Zhang2015PRB}%
  \BibitemOpen
  \bibfield  {author} {\bibinfo {author} {\bibfnamefont {B.}~\bibnamefont {Zhang}}, \bibinfo {author} {\bibfnamefont {C.-J.}\ \bibnamefont {Sun}}, \bibinfo {author} {\bibfnamefont {W.}~\bibnamefont {L\"u}}, \bibinfo {author} {\bibfnamefont {T.}~\bibnamefont {Venkatesan}}, \bibinfo {author} {\bibfnamefont {M.-G.}\ \bibnamefont {Han}}, \bibinfo {author} {\bibfnamefont {Y.}~\bibnamefont {Zhu}}, \bibinfo {author} {\bibfnamefont {J.}~\bibnamefont {Chen}},\ and\ \bibinfo {author} {\bibfnamefont {G.~M.}\ \bibnamefont {Chow}},\ }\href@noop {} {\bibfield  {journal} {\bibinfo  {journal} {Phys. Rev. B}\ }\textbf {\bibinfo {volume} {91}},\ \bibinfo {pages} {174431} (\bibinfo {year} {2015})}\BibitemShut {NoStop}%
\bibitem [{\citenamefont {Miwa}\ \emph {et~al.}(2015)\citenamefont {Miwa}, \citenamefont {Matsuda}, \citenamefont {Tanaka}, \citenamefont {Kotani}, \citenamefont {Goto}, \citenamefont {Nakamura},\ and\ \citenamefont {Suzuki}}]{Miwa2015APL}%
  \BibitemOpen
  \bibfield  {author} {\bibinfo {author} {\bibfnamefont {S.}~\bibnamefont {Miwa}}, \bibinfo {author} {\bibfnamefont {K.}~\bibnamefont {Matsuda}}, \bibinfo {author} {\bibfnamefont {K.}~\bibnamefont {Tanaka}}, \bibinfo {author} {\bibfnamefont {Y.}~\bibnamefont {Kotani}}, \bibinfo {author} {\bibfnamefont {M.}~\bibnamefont {Goto}}, \bibinfo {author} {\bibfnamefont {T.}~\bibnamefont {Nakamura}},\ and\ \bibinfo {author} {\bibfnamefont {Y.}~\bibnamefont {Suzuki}},\ }\href@noop {} {\bibfield  {journal} {\bibinfo  {journal} {Applied Physics Letters}\ }\textbf {\bibinfo {volume} {107}},\ \bibinfo {pages} {162402} (\bibinfo {year} {2015})}\BibitemShut {NoStop}%
\bibitem [{\citenamefont {Chang}\ \emph {et~al.}(2019)\citenamefont {Chang}, \citenamefont {Chung}, \citenamefont {Kao}, \citenamefont {Lee}, \citenamefont {Yu}, \citenamefont {Kaun}, \citenamefont {Nakamura}, \citenamefont {Sasabe}, \citenamefont {Chu},\ and\ \citenamefont {Tseng}}]{Chang20219ACSAPP}%
  \BibitemOpen
  \bibfield  {author} {\bibinfo {author} {\bibfnamefont {S.-J.}\ \bibnamefont {Chang}}, \bibinfo {author} {\bibfnamefont {M.-H.}\ \bibnamefont {Chung}}, \bibinfo {author} {\bibfnamefont {M.-Y.}\ \bibnamefont {Kao}}, \bibinfo {author} {\bibfnamefont {S.-F.}\ \bibnamefont {Lee}}, \bibinfo {author} {\bibfnamefont {Y.-H.}\ \bibnamefont {Yu}}, \bibinfo {author} {\bibfnamefont {C.-C.}\ \bibnamefont {Kaun}}, \bibinfo {author} {\bibfnamefont {T.}~\bibnamefont {Nakamura}}, \bibinfo {author} {\bibfnamefont {N.}~\bibnamefont {Sasabe}}, \bibinfo {author} {\bibfnamefont {S.-J.}\ \bibnamefont {Chu}},\ and\ \bibinfo {author} {\bibfnamefont {Y.-C.}\ \bibnamefont {Tseng}},\ }\href@noop {} {\bibfield  {journal} {\bibinfo  {journal} {ACS Applied Materials \& Interfaces}\ }\textbf {\bibinfo {volume} {11}},\ \bibinfo {pages} {31562} (\bibinfo {year} {2019})}\BibitemShut {NoStop}%
\bibitem [{\citenamefont {Ishii}\ \emph {et~al.}(2024)\citenamefont {Ishii}, \citenamefont {Yamasaki}, \citenamefont {Kozuka}, \citenamefont {Lustikova}, \citenamefont {Nii}, \citenamefont {Onose}, \citenamefont {Yokoyama}, \citenamefont {Mizumaki}, \citenamefont {Adachi}, \citenamefont {Nakao}, \citenamefont {Arima},\ and\ \citenamefont {Wakabayashi}}]{ishii2024microscopic}%
  \BibitemOpen
  \bibfield  {author} {\bibinfo {author} {\bibfnamefont {Y.}~\bibnamefont {Ishii}}, \bibinfo {author} {\bibfnamefont {Y.}~\bibnamefont {Yamasaki}}, \bibinfo {author} {\bibfnamefont {Y.}~\bibnamefont {Kozuka}}, \bibinfo {author} {\bibfnamefont {J.}~\bibnamefont {Lustikova}}, \bibinfo {author} {\bibfnamefont {Y.}~\bibnamefont {Nii}}, \bibinfo {author} {\bibfnamefont {Y.}~\bibnamefont {Onose}}, \bibinfo {author} {\bibfnamefont {Y.}~\bibnamefont {Yokoyama}}, \bibinfo {author} {\bibfnamefont {M.}~\bibnamefont {Mizumaki}}, \bibinfo {author} {\bibfnamefont {J.-i.}\ \bibnamefont {Adachi}}, \bibinfo {author} {\bibfnamefont {H.}~\bibnamefont {Nakao}}, \bibinfo {author} {\bibfnamefont {T.-h.}\ \bibnamefont {Arima}},\ and\ \bibinfo {author} {\bibfnamefont {Y.}~\bibnamefont {Wakabayashi}},\ }\href@noop {} {\bibfield  {journal} {\bibinfo  {journal} {Scientific reports}\ }\textbf {\bibinfo {volume} {14}},\ \bibinfo {pages} {15504} (\bibinfo {year} {2024})}\BibitemShut {NoStop}%
\bibitem [{\citenamefont {Chen}\ \emph {et~al.}(1997)\citenamefont {Chen}, \citenamefont {Sette}, \citenamefont {Ma},\ and\ \citenamefont {Modesti}}]{ChenXMLD1997}%
  \BibitemOpen
  \bibfield  {author} {\bibinfo {author} {\bibfnamefont {C.~T.}\ \bibnamefont {Chen}}, \bibinfo {author} {\bibfnamefont {F.}~\bibnamefont {Sette}}, \bibinfo {author} {\bibfnamefont {Y.}~\bibnamefont {Ma}},\ and\ \bibinfo {author} {\bibfnamefont {S.}~\bibnamefont {Modesti}},\ }\href@noop {} {\bibfield  {journal} {\bibinfo  {journal} {Phys. Rev. B}\ }\textbf {\bibinfo {volume} {56}},\ \bibinfo {pages} {R4959} (\bibinfo {year} {1997})}\BibitemShut {NoStop}%
\bibitem [{\citenamefont {Scholl}\ \emph {et~al.}(2000)\citenamefont {Scholl}, \citenamefont {Stohr}, \citenamefont {Luning}, \citenamefont {Seo}, \citenamefont {Fompeyrine}, \citenamefont {Siegwart}, \citenamefont {Locquet}, \citenamefont {Nolting}, \citenamefont {Anders}, \citenamefont {Fullerton} \emph {et~al.}}]{scholl2000observation}%
  \BibitemOpen
  \bibfield  {author} {\bibinfo {author} {\bibfnamefont {A.}~\bibnamefont {Scholl}}, \bibinfo {author} {\bibfnamefont {J.}~\bibnamefont {Stohr}}, \bibinfo {author} {\bibfnamefont {J.}~\bibnamefont {Luning}}, \bibinfo {author} {\bibfnamefont {J.~W.}\ \bibnamefont {Seo}}, \bibinfo {author} {\bibfnamefont {J.}~\bibnamefont {Fompeyrine}}, \bibinfo {author} {\bibfnamefont {H.}~\bibnamefont {Siegwart}}, \bibinfo {author} {\bibfnamefont {J.-P.}\ \bibnamefont {Locquet}}, \bibinfo {author} {\bibfnamefont {F.}~\bibnamefont {Nolting}}, \bibinfo {author} {\bibfnamefont {S.}~\bibnamefont {Anders}}, \bibinfo {author} {\bibfnamefont {E.}~\bibnamefont {Fullerton}}, \emph {et~al.},\ }\href@noop {} {\bibfield  {journal} {\bibinfo  {journal} {Science}\ }\textbf {\bibinfo {volume} {287}},\ \bibinfo {pages} {1014} (\bibinfo {year} {2000})}\BibitemShut {NoStop}%
\bibitem [{\citenamefont {Nakatsuji}\ \emph {et~al.}(2015)\citenamefont {Nakatsuji}, \citenamefont {Kiyohara},\ and\ \citenamefont {Higo}}]{Nakatsuji2015}%
  \BibitemOpen
  \bibfield  {author} {\bibinfo {author} {\bibfnamefont {S.}~\bibnamefont {Nakatsuji}}, \bibinfo {author} {\bibfnamefont {N.}~\bibnamefont {Kiyohara}},\ and\ \bibinfo {author} {\bibfnamefont {T.}~\bibnamefont {Higo}},\ }\href@noop {} {\bibfield  {journal} {\bibinfo  {journal} {Nature}\ }\textbf {\bibinfo {volume} {527}},\ \bibinfo {pages} {212} (\bibinfo {year} {2015})}\BibitemShut {NoStop}%
\bibitem [{\citenamefont {Higo}\ \emph {et~al.}(2018)\citenamefont {Higo}, \citenamefont {Man}, \citenamefont {Gopman}, \citenamefont {Wu}, \citenamefont {Koretsune}, \citenamefont {van’t Erve}, \citenamefont {Kabanov}, \citenamefont {Rees}, \citenamefont {Li}, \citenamefont {Suzuki} \emph {et~al.}}]{Mn3Sn_2018_MOKE_higo}%
  \BibitemOpen
  \bibfield  {author} {\bibinfo {author} {\bibfnamefont {T.}~\bibnamefont {Higo}}, \bibinfo {author} {\bibfnamefont {H.}~\bibnamefont {Man}}, \bibinfo {author} {\bibfnamefont {D.~B.}\ \bibnamefont {Gopman}}, \bibinfo {author} {\bibfnamefont {L.}~\bibnamefont {Wu}}, \bibinfo {author} {\bibfnamefont {T.}~\bibnamefont {Koretsune}}, \bibinfo {author} {\bibfnamefont {O.~M.}\ \bibnamefont {van’t Erve}}, \bibinfo {author} {\bibfnamefont {Y.~P.}\ \bibnamefont {Kabanov}}, \bibinfo {author} {\bibfnamefont {D.}~\bibnamefont {Rees}}, \bibinfo {author} {\bibfnamefont {Y.}~\bibnamefont {Li}}, \bibinfo {author} {\bibfnamefont {M.-T.}\ \bibnamefont {Suzuki}}, \emph {et~al.},\ }\href@noop {} {\bibfield  {journal} {\bibinfo  {journal} {Nature photonics}\ }\textbf {\bibinfo {volume} {12}},\ \bibinfo {pages} {73} (\bibinfo {year} {2018})}\BibitemShut {NoStop}%
\bibitem [{\citenamefont {Solovyev}(1997)}]{Solovyev1997}%
  \BibitemOpen
  \bibfield  {author} {\bibinfo {author} {\bibfnamefont {I.~V.}\ \bibnamefont {Solovyev}},\ }\href@noop {} {\bibfield  {journal} {\bibinfo  {journal} {Phys. Rev. B}\ }\textbf {\bibinfo {volume} {55}},\ \bibinfo {pages} {8060} (\bibinfo {year} {1997})}\BibitemShut {NoStop}%
\bibitem [{\citenamefont {Noda}\ \emph {et~al.}(2016)\citenamefont {Noda}, \citenamefont {Ohno},\ and\ \citenamefont {Nakamura}}]{Noda2016_PhysChemChemPhys}%
  \BibitemOpen
  \bibfield  {author} {\bibinfo {author} {\bibfnamefont {Y.}~\bibnamefont {Noda}}, \bibinfo {author} {\bibfnamefont {K.}~\bibnamefont {Ohno}},\ and\ \bibinfo {author} {\bibfnamefont {S.}~\bibnamefont {Nakamura}},\ }\href@noop {} {\bibfield  {journal} {\bibinfo  {journal} {Phys. Chem. Chem. Phys.}\ }\textbf {\bibinfo {volume} {18}},\ \bibinfo {pages} {13294} (\bibinfo {year} {2016})}\BibitemShut {NoStop}%
\bibitem [{\citenamefont {Okugawa}\ \emph {et~al.}(2018)\citenamefont {Okugawa}, \citenamefont {Ohno}, \citenamefont {Noda},\ and\ \citenamefont {Nakamura}}]{Okugawa_AM2018}%
  \BibitemOpen
  \bibfield  {author} {\bibinfo {author} {\bibfnamefont {T.}~\bibnamefont {Okugawa}}, \bibinfo {author} {\bibfnamefont {K.}~\bibnamefont {Ohno}}, \bibinfo {author} {\bibfnamefont {Y.}~\bibnamefont {Noda}},\ and\ \bibinfo {author} {\bibfnamefont {S.}~\bibnamefont {Nakamura}},\ }\href@noop {} {\bibfield  {journal} {\bibinfo  {journal} {Journal of Physics: Condensed Matter}\ }\textbf {\bibinfo {volume} {30}},\ \bibinfo {pages} {075502} (\bibinfo {year} {2018})}\BibitemShut {NoStop}%
\bibitem [{\citenamefont {Naka}\ \emph {et~al.}(2019)\citenamefont {Naka}, \citenamefont {Hayami}, \citenamefont {Kusunose}, \citenamefont {Yanagi}, \citenamefont {Motome},\ and\ \citenamefont {Seo}}]{Naka2019NatCom}%
  \BibitemOpen
  \bibfield  {author} {\bibinfo {author} {\bibfnamefont {M.}~\bibnamefont {Naka}}, \bibinfo {author} {\bibfnamefont {S.}~\bibnamefont {Hayami}}, \bibinfo {author} {\bibfnamefont {H.}~\bibnamefont {Kusunose}}, \bibinfo {author} {\bibfnamefont {Y.}~\bibnamefont {Yanagi}}, \bibinfo {author} {\bibfnamefont {Y.}~\bibnamefont {Motome}},\ and\ \bibinfo {author} {\bibfnamefont {H.}~\bibnamefont {Seo}},\ }\href@noop {} {\bibfield  {journal} {\bibinfo  {journal} {Nature Communications}\ }\textbf {\bibinfo {volume} {10}} (\bibinfo {year} {2019})}\BibitemShut {NoStop}%
\bibitem [{\citenamefont {Ahn}\ \emph {et~al.}(2019)\citenamefont {Ahn}, \citenamefont {Hariki}, \citenamefont {Lee},\ and\ \citenamefont {Kune\ifmmode~\check{s}\else \v{s}\fi{}}}]{Ahn2019PRB}%
  \BibitemOpen
  \bibfield  {author} {\bibinfo {author} {\bibfnamefont {K.-H.}\ \bibnamefont {Ahn}}, \bibinfo {author} {\bibfnamefont {A.}~\bibnamefont {Hariki}}, \bibinfo {author} {\bibfnamefont {K.-W.}\ \bibnamefont {Lee}},\ and\ \bibinfo {author} {\bibfnamefont {J.}~\bibnamefont {Kune\ifmmode~\check{s}\else \v{s}\fi{}}},\ }\href@noop {} {\bibfield  {journal} {\bibinfo  {journal} {Phys. Rev. B}\ }\textbf {\bibinfo {volume} {99}},\ \bibinfo {pages} {184432} (\bibinfo {year} {2019})}\BibitemShut {NoStop}%
\bibitem [{\citenamefont {\v{S}mejkal}\ \emph {et~al.}(2020)\citenamefont {\v{S}mejkal}, \citenamefont {Gonzalez-Hernandez}, \citenamefont {Jungwirth},\ and\ \citenamefont {Sinova}}]{Smejkal2020_SciAdv}%
  \BibitemOpen
  \bibfield  {author} {\bibinfo {author} {\bibfnamefont {L.}~\bibnamefont {\v{S}mejkal}}, \bibinfo {author} {\bibfnamefont {R.}~\bibnamefont {Gonzalez-Hernandez}}, \bibinfo {author} {\bibfnamefont {T.}~\bibnamefont {Jungwirth}},\ and\ \bibinfo {author} {\bibfnamefont {J.}~\bibnamefont {Sinova}},\ }\href@noop {} {\bibfield  {journal} {\bibinfo  {journal} {Sci. Adv.}\ }\textbf {\bibinfo {volume} {6}},\ \bibinfo {pages} {eaba9399} (\bibinfo {year} {2020})}\BibitemShut {NoStop}%
\bibitem [{\citenamefont {\v{S}mejkal}\ \emph {et~al.}(2022)\citenamefont {\v{S}mejkal}, \citenamefont {Gonzalez-Hernandez}, \citenamefont {Jungwirth},\ and\ \citenamefont {Sinova}}]{Smejkal2022}%
  \BibitemOpen
  \bibfield  {author} {\bibinfo {author} {\bibfnamefont {L.}~\bibnamefont {\v{S}mejkal}}, \bibinfo {author} {\bibfnamefont {R.}~\bibnamefont {Gonzalez-Hernandez}}, \bibinfo {author} {\bibfnamefont {T.}~\bibnamefont {Jungwirth}},\ and\ \bibinfo {author} {\bibfnamefont {J.}~\bibnamefont {Sinova}},\ }\href@noop {} {\bibfield  {journal} {\bibinfo  {journal} {Science}\ }\textbf {\bibinfo {volume} {375}},\ \bibinfo {pages} {70} (\bibinfo {year} {2022})}\BibitemShut {NoStop}%
\bibitem [{\citenamefont {\v{S}mejkal}\ \emph {et~al.}(2023)\citenamefont {\v{S}mejkal}, \citenamefont {Jungwirth},\ and\ \citenamefont {Sinova}}]{Smejkal2023_NatRevMater}%
  \BibitemOpen
  \bibfield  {author} {\bibinfo {author} {\bibfnamefont {L.}~\bibnamefont {\v{S}mejkal}}, \bibinfo {author} {\bibfnamefont {T.}~\bibnamefont {Jungwirth}},\ and\ \bibinfo {author} {\bibfnamefont {J.}~\bibnamefont {Sinova}},\ }\href@noop {} {\bibfield  {journal} {\bibinfo  {journal} {Nat. Rev. Mater.}\ }\textbf {\bibinfo {volume} {8}},\ \bibinfo {pages} {653} (\bibinfo {year} {2023})}\BibitemShut {NoStop}%
\bibitem [{\citenamefont {Hariki}\ \emph {et~al.}(2024)\citenamefont {Hariki}, \citenamefont {Dal~Din}, \citenamefont {Amin}, \citenamefont {Yamaguchi}, \citenamefont {Badura}, \citenamefont {Kriegner}, \citenamefont {Edmonds}, \citenamefont {Campion}, \citenamefont {Wadley}, \citenamefont {Backes}, \citenamefont {Veiga}, \citenamefont {Dhesi}, \citenamefont {Springholz}, \citenamefont {\ifmmode~\check{S}\else \v{S}\fi{}mejkal}, \citenamefont {V\'yborn\'y}, \citenamefont {Jungwirth},\ and\ \citenamefont {Kune\ifmmode~\check{s}\else \v{s}\fi{}}}]{Hariki2024PRL}%
  \BibitemOpen
  \bibfield  {author} {\bibinfo {author} {\bibfnamefont {A.}~\bibnamefont {Hariki}}, \bibinfo {author} {\bibfnamefont {A.}~\bibnamefont {Dal~Din}}, \bibinfo {author} {\bibfnamefont {O.~J.}\ \bibnamefont {Amin}}, \bibinfo {author} {\bibfnamefont {T.}~\bibnamefont {Yamaguchi}}, \bibinfo {author} {\bibfnamefont {A.}~\bibnamefont {Badura}}, \bibinfo {author} {\bibfnamefont {D.}~\bibnamefont {Kriegner}}, \bibinfo {author} {\bibfnamefont {K.~W.}\ \bibnamefont {Edmonds}}, \bibinfo {author} {\bibfnamefont {R.~P.}\ \bibnamefont {Campion}}, \bibinfo {author} {\bibfnamefont {P.}~\bibnamefont {Wadley}}, \bibinfo {author} {\bibfnamefont {D.}~\bibnamefont {Backes}}, \bibinfo {author} {\bibfnamefont {L.~S.~I.}\ \bibnamefont {Veiga}}, \bibinfo {author} {\bibfnamefont {S.~S.}\ \bibnamefont {Dhesi}}, \bibinfo {author} {\bibfnamefont {G.}~\bibnamefont {Springholz}}, \bibinfo {author} {\bibfnamefont {L.}~\bibnamefont {\ifmmode~\check{S}\else \v{S}\fi{}mejkal}}, \bibinfo {author} {\bibfnamefont {K.}~\bibnamefont {V\'yborn\'y}},
  \bibinfo {author} {\bibfnamefont {T.}~\bibnamefont {Jungwirth}},\ and\ \bibinfo {author} {\bibfnamefont {J.}~\bibnamefont {Kune\ifmmode~\check{s}\else \v{s}\fi{}}},\ }\href@noop {} {\bibfield  {journal} {\bibinfo  {journal} {Phys. Rev. Lett.}\ }\textbf {\bibinfo {volume} {132}},\ \bibinfo {pages} {176701} (\bibinfo {year} {2024})}\BibitemShut {NoStop}%
\bibitem [{\citenamefont {Amin}\ \emph {et~al.}(2024{\natexlab{a}})\citenamefont {Amin}, \citenamefont {Din}, \citenamefont {Golias}, \citenamefont {Niu}, \citenamefont {Zakharov}, \citenamefont {Fromage}, \citenamefont {Fields}, \citenamefont {Heywood}, \citenamefont {Cousins}, \citenamefont {Maccherozzi}, \citenamefont {Krempask\'{y}}, \citenamefont {Dil}, \citenamefont {Kriegner}, \citenamefont {Kiraly}, \citenamefont {Campion}, \citenamefont {Rushforth}, \citenamefont {Edmonds}, \citenamefont {Dhesi}, \citenamefont {\v{S}mejkal}, \citenamefont {Jungwirth},\ and\ \citenamefont {Wadley}}]{MnTe2024Nature}%
  \BibitemOpen
  \bibfield  {author} {\bibinfo {author} {\bibfnamefont {O.~J.}\ \bibnamefont {Amin}}, \bibinfo {author} {\bibfnamefont {A.~D.}\ \bibnamefont {Din}}, \bibinfo {author} {\bibfnamefont {E.}~\bibnamefont {Golias}}, \bibinfo {author} {\bibfnamefont {Y.}~\bibnamefont {Niu}}, \bibinfo {author} {\bibfnamefont {A.}~\bibnamefont {Zakharov}}, \bibinfo {author} {\bibfnamefont {S.~C.}\ \bibnamefont {Fromage}}, \bibinfo {author} {\bibfnamefont {C.~J.~B.}\ \bibnamefont {Fields}}, \bibinfo {author} {\bibfnamefont {S.~L.}\ \bibnamefont {Heywood}}, \bibinfo {author} {\bibfnamefont {R.~B.}\ \bibnamefont {Cousins}}, \bibinfo {author} {\bibfnamefont {F.}~\bibnamefont {Maccherozzi}}, \bibinfo {author} {\bibfnamefont {J.}~\bibnamefont {Krempask\'{y}}}, \bibinfo {author} {\bibfnamefont {J.~H.}\ \bibnamefont {Dil}}, \bibinfo {author} {\bibfnamefont {D.}~\bibnamefont {Kriegner}}, \bibinfo {author} {\bibfnamefont {B.}~\bibnamefont {Kiraly}}, \bibinfo {author} {\bibfnamefont {R.~P.}\ \bibnamefont {Campion}}, \bibinfo {author}
  {\bibfnamefont {A.~W.}\ \bibnamefont {Rushforth}}, \bibinfo {author} {\bibfnamefont {K.~W.}\ \bibnamefont {Edmonds}}, \bibinfo {author} {\bibfnamefont {S.~S.}\ \bibnamefont {Dhesi}}, \bibinfo {author} {\bibfnamefont {L.}~\bibnamefont {\v{S}mejkal}}, \bibinfo {author} {\bibfnamefont {T.}~\bibnamefont {Jungwirth}},\ and\ \bibinfo {author} {\bibfnamefont {P.}~\bibnamefont {Wadley}},\ }\href@noop {} {\bibfield  {journal} {\bibinfo  {journal} {Nature}\ }\textbf {\bibinfo {volume} {636}},\ \bibinfo {pages} {348} (\bibinfo {year} {2024}{\natexlab{a}})}\BibitemShut {NoStop}%
\bibitem [{\citenamefont {Yamasaki}\ \emph {et~al.}(2020)\citenamefont {Yamasaki}, \citenamefont {Nakao},\ and\ \citenamefont {Arima}}]{Yamasaki2020}%
  \BibitemOpen
  \bibfield  {author} {\bibinfo {author} {\bibfnamefont {Y.}~\bibnamefont {Yamasaki}}, \bibinfo {author} {\bibfnamefont {H.}~\bibnamefont {Nakao}},\ and\ \bibinfo {author} {\bibfnamefont {T.-h.}\ \bibnamefont {Arima}},\ }\href@noop {} {\bibfield  {journal} {\bibinfo  {journal} {Journal of the Physical Society of Japan}\ }\textbf {\bibinfo {volume} {89}},\ \bibinfo {pages} {083703} (\bibinfo {year} {2020})}\BibitemShut {NoStop}%
\bibitem [{\citenamefont {Sasabe}\ \emph {et~al.}(2021)\citenamefont {Sasabe}, \citenamefont {Kimata},\ and\ \citenamefont {Nakamura}}]{Sasabe_PRL_Mn3Sn}%
  \BibitemOpen
  \bibfield  {author} {\bibinfo {author} {\bibfnamefont {N.}~\bibnamefont {Sasabe}}, \bibinfo {author} {\bibfnamefont {M.}~\bibnamefont {Kimata}},\ and\ \bibinfo {author} {\bibfnamefont {T.}~\bibnamefont {Nakamura}},\ }\href@noop {} {\bibfield  {journal} {\bibinfo  {journal} {Phys. Rev. Lett.}\ }\textbf {\bibinfo {volume} {126}},\ \bibinfo {pages} {157402} (\bibinfo {year} {2021})}\BibitemShut {NoStop}%
\bibitem [{\citenamefont {Kimata}\ \emph {et~al.}(2021)\citenamefont {Kimata}, \citenamefont {Sasabe}, \citenamefont {Kurita}, \citenamefont {Yamasaki}, \citenamefont {Tabata}, \citenamefont {Yokoyama}, \citenamefont {Kotani}, \citenamefont {Ikhlas}, \citenamefont {Tomita}, \citenamefont {Amemiya}, \citenamefont {Nojiri}, \citenamefont {Nakatsuji}, \citenamefont {Koretsune}, \citenamefont {Nakao}, \citenamefont {Arima},\ and\ \citenamefont {Nakamura}}]{XMCD_Kimata}%
  \BibitemOpen
  \bibfield  {author} {\bibinfo {author} {\bibfnamefont {M.}~\bibnamefont {Kimata}}, \bibinfo {author} {\bibfnamefont {N.}~\bibnamefont {Sasabe}}, \bibinfo {author} {\bibfnamefont {K.}~\bibnamefont {Kurita}}, \bibinfo {author} {\bibfnamefont {Y.}~\bibnamefont {Yamasaki}}, \bibinfo {author} {\bibfnamefont {C.}~\bibnamefont {Tabata}}, \bibinfo {author} {\bibfnamefont {Y.}~\bibnamefont {Yokoyama}}, \bibinfo {author} {\bibfnamefont {Y.}~\bibnamefont {Kotani}}, \bibinfo {author} {\bibfnamefont {M.}~\bibnamefont {Ikhlas}}, \bibinfo {author} {\bibfnamefont {T.}~\bibnamefont {Tomita}}, \bibinfo {author} {\bibfnamefont {K.}~\bibnamefont {Amemiya}}, \bibinfo {author} {\bibfnamefont {H.}~\bibnamefont {Nojiri}}, \bibinfo {author} {\bibfnamefont {S.}~\bibnamefont {Nakatsuji}}, \bibinfo {author} {\bibfnamefont {T.}~\bibnamefont {Koretsune}}, \bibinfo {author} {\bibfnamefont {H.}~\bibnamefont {Nakao}}, \bibinfo {author} {\bibfnamefont {T.-h.}\ \bibnamefont {Arima}},\ and\ \bibinfo {author} {\bibfnamefont {T.}~\bibnamefont
  {Nakamura}},\ }\href@noop {} {\bibfield  {journal} {\bibinfo  {journal} {Nat. Commu.}\ }\textbf {\bibinfo {volume} {12}},\ \bibinfo {pages} {5582} (\bibinfo {year} {2021})}\BibitemShut {NoStop}%
\bibitem [{\citenamefont {Sakamoto}\ \emph {et~al.}(2021)\citenamefont {Sakamoto}, \citenamefont {Higo}, \citenamefont {Shiga}, \citenamefont {Amemiya}, \citenamefont {Nakatsuji},\ and\ \citenamefont {Miwa}}]{sakamoto2021observation}%
  \BibitemOpen
  \bibfield  {author} {\bibinfo {author} {\bibfnamefont {S.}~\bibnamefont {Sakamoto}}, \bibinfo {author} {\bibfnamefont {T.}~\bibnamefont {Higo}}, \bibinfo {author} {\bibfnamefont {M.}~\bibnamefont {Shiga}}, \bibinfo {author} {\bibfnamefont {K.}~\bibnamefont {Amemiya}}, \bibinfo {author} {\bibfnamefont {S.}~\bibnamefont {Nakatsuji}},\ and\ \bibinfo {author} {\bibfnamefont {S.}~\bibnamefont {Miwa}},\ }\href@noop {} {\bibfield  {journal} {\bibinfo  {journal} {Physical Review B}\ }\textbf {\bibinfo {volume} {104}},\ \bibinfo {pages} {134431} (\bibinfo {year} {2021})}\BibitemShut {NoStop}%
\bibitem [{\citenamefont {Sakamoto}\ \emph {et~al.}(2024)\citenamefont {Sakamoto}, \citenamefont {Higo}, \citenamefont {Kotani}, \citenamefont {Kosaki}, \citenamefont {Nakamura}, \citenamefont {Nakatsuji},\ and\ \citenamefont {Miwa}}]{sakamoto2024bulk}%
  \BibitemOpen
  \bibfield  {author} {\bibinfo {author} {\bibfnamefont {S.}~\bibnamefont {Sakamoto}}, \bibinfo {author} {\bibfnamefont {T.}~\bibnamefont {Higo}}, \bibinfo {author} {\bibfnamefont {Y.}~\bibnamefont {Kotani}}, \bibinfo {author} {\bibfnamefont {H.}~\bibnamefont {Kosaki}}, \bibinfo {author} {\bibfnamefont {T.}~\bibnamefont {Nakamura}}, \bibinfo {author} {\bibfnamefont {S.}~\bibnamefont {Nakatsuji}},\ and\ \bibinfo {author} {\bibfnamefont {S.}~\bibnamefont {Miwa}},\ }\href@noop {} {\bibfield  {journal} {\bibinfo  {journal} {Physical Review B}\ }\textbf {\bibinfo {volume} {110}},\ \bibinfo {pages} {L060412} (\bibinfo {year} {2024})}\BibitemShut {NoStop}%
\bibitem [{\citenamefont {Kusunose}\ \emph {et~al.}(2020)\citenamefont {Kusunose}, \citenamefont {Oiwa},\ and\ \citenamefont {Hayami}}]{kusunose2020complete}%
  \BibitemOpen
  \bibfield  {author} {\bibinfo {author} {\bibfnamefont {H.}~\bibnamefont {Kusunose}}, \bibinfo {author} {\bibfnamefont {R.}~\bibnamefont {Oiwa}},\ and\ \bibinfo {author} {\bibfnamefont {S.}~\bibnamefont {Hayami}},\ }\href@noop {} {\bibfield  {journal} {\bibinfo  {journal} {Journal of the Physical Society of Japan}\ }\textbf {\bibinfo {volume} {89}},\ \bibinfo {pages} {104704} (\bibinfo {year} {2020})}\BibitemShut {NoStop}%
\bibitem [{\citenamefont {Hayami}\ and\ \citenamefont {Kusunose}(2024)}]{hayami2024unified}%
  \BibitemOpen
  \bibfield  {author} {\bibinfo {author} {\bibfnamefont {S.}~\bibnamefont {Hayami}}\ and\ \bibinfo {author} {\bibfnamefont {H.}~\bibnamefont {Kusunose}},\ }\href@noop {} {\bibfield  {journal} {\bibinfo  {journal} {Journal of the Physical Society of Japan}\ }\textbf {\bibinfo {volume} {93}},\ \bibinfo {pages} {072001} (\bibinfo {year} {2024})}\BibitemShut {NoStop}%
\bibitem [{Here, we use the following relation, $ \sum_{m_j}C_{l_cm_c;\frac{1}{2}m_s}^{jm_j} C_{l_cm_c^\prime;\frac{1}{2}m_s^\prime}^{jm_j} = \frac{j_\pm+\frac{1}{2}}{2l_c+1} C_{\frac{1}{2}m_s^\prime;00}^{\frac{1}{2}m_s} C_{l_cm_c^\prime;00}^{l_cm_c} \pm \sum_{n=-1}^1 \frac{\sqrt{3l_c(l_c+1)}}{2l_c+1} C_{\frac{1}{2}m_s^\prime;1n}^{\frac{1}{2}m_s} C_{l_cm_c^\prime;1,-n}^{l_cm_c}$()}]{formula1}%
  \BibitemOpen
  Here, we use the following relation, $ \sum_{m_j}C_{l_cm_c;\frac{1}{2}m_s}^{jm_j} C_{l_cm_c^\prime;\frac{1}{2}m_s^\prime}^{jm_j} = \frac{j_\pm+\frac{1}{2}}{2l_c+1} C_{\frac{1}{2}m_s^\prime;00}^{\frac{1}{2}m_s} C_{l_cm_c^\prime;00}^{l_cm_c} \pm \sum_{n=-1}^1 \frac{\sqrt{3l_c(l_c+1)}}{2l_c+1} C_{\frac{1}{2}m_s^\prime;1n}^{\frac{1}{2}m_s} C_{l_cm_c^\prime;1,-n}^{l_cm_c}$,\ \href@noop {} {}\BibitemShut {NoStop}%
\bibitem [{for()}]{formulra3}%
  \BibitemOpen
  \bibinfo {note} {The function for finite summation involving four Clebsh-Gordan coefficients used here is found in https://functions.wolfram.com/07.38.23.0029.01}\BibitemShut {NoStop}%
\bibitem [{\citenamefont {Carra}\ \emph {et~al.}(1993{\natexlab{b}})\citenamefont {Carra}, \citenamefont {K{\"o}nig}, \citenamefont {Thole},\ and\ \citenamefont {Altarelli}}]{carra1993magnetic}%
  \BibitemOpen
  \bibfield  {author} {\bibinfo {author} {\bibfnamefont {P.}~\bibnamefont {Carra}}, \bibinfo {author} {\bibfnamefont {H.}~\bibnamefont {K{\"o}nig}}, \bibinfo {author} {\bibfnamefont {B.}~\bibnamefont {Thole}},\ and\ \bibinfo {author} {\bibfnamefont {M.}~\bibnamefont {Altarelli}},\ }\href@noop {} {\bibfield  {journal} {\bibinfo  {journal} {Physica B: Condensed Matter}\ }\textbf {\bibinfo {volume} {192}},\ \bibinfo {pages} {182} (\bibinfo {year} {1993}{\natexlab{b}})}\BibitemShut {NoStop}%
\bibitem [{\citenamefont {Ankudinov}\ \emph {et~al.}(2003)\citenamefont {Ankudinov}, \citenamefont {Nesvizhskii},\ and\ \citenamefont {Rehr}}]{Ankudinov2003-vd}%
  \BibitemOpen
  \bibfield  {author} {\bibinfo {author} {\bibfnamefont {A.~L.}\ \bibnamefont {Ankudinov}}, \bibinfo {author} {\bibfnamefont {A.~I.}\ \bibnamefont {Nesvizhskii}},\ and\ \bibinfo {author} {\bibfnamefont {J.~J.}\ \bibnamefont {Rehr}},\ }\href@noop {} {\bibfield  {journal} {\bibinfo  {journal} {Phys. Rev. B Condens. Matter}\ }\textbf {\bibinfo {volume} {67}},\ \bibinfo {pages} {115120} (\bibinfo {year} {2003})}\BibitemShut {NoStop}%
\bibitem [{\citenamefont {Schwitalla}\ and\ \citenamefont {Ebert}(1998)}]{Schwitalla1998-ro}%
  \BibitemOpen
  \bibfield  {author} {\bibinfo {author} {\bibfnamefont {J.}~\bibnamefont {Schwitalla}}\ and\ \bibinfo {author} {\bibfnamefont {H.}~\bibnamefont {Ebert}},\ }\href@noop {} {\bibfield  {journal} {\bibinfo  {journal} {Phys. Rev. Lett.}\ }\textbf {\bibinfo {volume} {80}},\ \bibinfo {pages} {4586} (\bibinfo {year} {1998})}\BibitemShut {NoStop}%
\bibitem [{\citenamefont {{T. Oguchi }}\ and\ \citenamefont {{T. Shishidou}}(2004)}]{Oguchi}%
  \BibitemOpen
  \bibfield  {author} {\bibinfo {author} {\bibnamefont {{T. Oguchi }}}\ and\ \bibinfo {author} {\bibnamefont {{T. Shishidou}}},\ }\href@noop {} {\bibfield  {journal} {\bibinfo  {journal} {Phys. Rev. B}\ }\textbf {\bibinfo {volume} {70}},\ \bibinfo {pages} {024412} (\bibinfo {year} {2004})}\BibitemShut {NoStop}%
\bibitem [{\citenamefont {van~der Laan}(1998)}]{PhysRevB.57.112}%
  \BibitemOpen
  \bibfield  {author} {\bibinfo {author} {\bibfnamefont {G.}~\bibnamefont {van~der Laan}},\ }\href@noop {} {\bibfield  {journal} {\bibinfo  {journal} {Phys. Rev. B}\ }\textbf {\bibinfo {volume} {57}},\ \bibinfo {pages} {112} (\bibinfo {year} {1998})}\BibitemShut {NoStop}%
\bibitem [{\citenamefont {Bitla}\ \emph {et~al.}(2015)\citenamefont {Bitla}, \citenamefont {Chin}, \citenamefont {Lin}, \citenamefont {Van}, \citenamefont {Liu}, \citenamefont {Zhu}, \citenamefont {Liu}, \citenamefont {Zhan}, \citenamefont {Lin}, \citenamefont {Chen}, \citenamefont {Chu},\ and\ \citenamefont {He}}]{Bitla2015-hs}%
  \BibitemOpen
  \bibfield  {author} {\bibinfo {author} {\bibfnamefont {Y.}~\bibnamefont {Bitla}}, \bibinfo {author} {\bibfnamefont {Y.-Y.}\ \bibnamefont {Chin}}, \bibinfo {author} {\bibfnamefont {J.-C.}\ \bibnamefont {Lin}}, \bibinfo {author} {\bibfnamefont {C.~N.}\ \bibnamefont {Van}}, \bibinfo {author} {\bibfnamefont {R.}~\bibnamefont {Liu}}, \bibinfo {author} {\bibfnamefont {Y.}~\bibnamefont {Zhu}}, \bibinfo {author} {\bibfnamefont {H.-J.}\ \bibnamefont {Liu}}, \bibinfo {author} {\bibfnamefont {Q.}~\bibnamefont {Zhan}}, \bibinfo {author} {\bibfnamefont {H.-J.}\ \bibnamefont {Lin}}, \bibinfo {author} {\bibfnamefont {C.-T.}\ \bibnamefont {Chen}}, \bibinfo {author} {\bibfnamefont {Y.-H.}\ \bibnamefont {Chu}},\ and\ \bibinfo {author} {\bibfnamefont {Q.}~\bibnamefont {He}},\ }\href@noop {} {\bibfield  {journal} {\bibinfo  {journal} {Sci. Rep.}\ }\textbf {\bibinfo {volume} {5}},\ \bibinfo {pages} {15201} (\bibinfo {year} {2015})}\BibitemShut {NoStop}%
\bibitem [{\citenamefont {Sasabe}\ \emph {et~al.}(2023)\citenamefont {Sasabe}, \citenamefont {Mizumaki}, \citenamefont {Uozumi},\ and\ \citenamefont {Yamasaki}}]{sasabe2023ferroic}%
  \BibitemOpen
  \bibfield  {author} {\bibinfo {author} {\bibfnamefont {N.}~\bibnamefont {Sasabe}}, \bibinfo {author} {\bibfnamefont {M.}~\bibnamefont {Mizumaki}}, \bibinfo {author} {\bibfnamefont {T.}~\bibnamefont {Uozumi}},\ and\ \bibinfo {author} {\bibfnamefont {Y.}~\bibnamefont {Yamasaki}},\ }\href@noop {} {\bibfield  {journal} {\bibinfo  {journal} {Physical Review Letters}\ }\textbf {\bibinfo {volume} {131}},\ \bibinfo {pages} {216501} (\bibinfo {year} {2023})}\BibitemShut {NoStop}%
\bibitem [{\citenamefont {Amin}\ \emph {et~al.}(2024{\natexlab{b}})\citenamefont {Amin}, \citenamefont {Dal~Din}, \citenamefont {Golias}, \citenamefont {Niu}, \citenamefont {Zakharov}, \citenamefont {Fromage}, \citenamefont {Fields}, \citenamefont {Heywood}, \citenamefont {Cousins}, \citenamefont {Maccherozzi} \emph {et~al.}}]{amin2024nanoscale}%
  \BibitemOpen
  \bibfield  {author} {\bibinfo {author} {\bibfnamefont {O.}~\bibnamefont {Amin}}, \bibinfo {author} {\bibfnamefont {A.}~\bibnamefont {Dal~Din}}, \bibinfo {author} {\bibfnamefont {E.}~\bibnamefont {Golias}}, \bibinfo {author} {\bibfnamefont {Y.}~\bibnamefont {Niu}}, \bibinfo {author} {\bibfnamefont {A.}~\bibnamefont {Zakharov}}, \bibinfo {author} {\bibfnamefont {S.}~\bibnamefont {Fromage}}, \bibinfo {author} {\bibfnamefont {C.}~\bibnamefont {Fields}}, \bibinfo {author} {\bibfnamefont {S.}~\bibnamefont {Heywood}}, \bibinfo {author} {\bibfnamefont {R.}~\bibnamefont {Cousins}}, \bibinfo {author} {\bibfnamefont {F.}~\bibnamefont {Maccherozzi}}, \emph {et~al.},\ }\href@noop {} {\bibfield  {journal} {\bibinfo  {journal} {Nature}\ }\textbf {\bibinfo {volume} {636}},\ \bibinfo {pages} {348} (\bibinfo {year} {2024}{\natexlab{b}})}\BibitemShut {NoStop}%
\bibitem [{\citenamefont {Hayami}\ and\ \citenamefont {Kusunose}(2021)}]{hayami2021essential}%
  \BibitemOpen
  \bibfield  {author} {\bibinfo {author} {\bibfnamefont {S.}~\bibnamefont {Hayami}}\ and\ \bibinfo {author} {\bibfnamefont {H.}~\bibnamefont {Kusunose}},\ }\href@noop {} {\bibfield  {journal} {\bibinfo  {journal} {Phys. Rev. B}\ }\textbf {\bibinfo {volume} {103}},\ \bibinfo {pages} {L180407} (\bibinfo {year} {2021})}\BibitemShut {NoStop}%
\bibitem [{\citenamefont {van~der Laan}(1999)}]{van1999magnetic}%
  \BibitemOpen
  \bibfield  {author} {\bibinfo {author} {\bibfnamefont {G.}~\bibnamefont {van~der Laan}},\ }\href@noop {} {\bibfield  {journal} {\bibinfo  {journal} {Physical review letters}\ }\textbf {\bibinfo {volume} {82}},\ \bibinfo {pages} {640} (\bibinfo {year} {1999})}\BibitemShut {NoStop}%
\bibitem [{\citenamefont {Okabayashi}\ \emph {et~al.}(2014)\citenamefont {Okabayashi}, \citenamefont {Koo}, \citenamefont {Sukegawa}, \citenamefont {Mitani}, \citenamefont {Takagi},\ and\ \citenamefont {Yokoyama}}]{Okabayashi2014-rp}%
  \BibitemOpen
  \bibfield  {author} {\bibinfo {author} {\bibfnamefont {J.}~\bibnamefont {Okabayashi}}, \bibinfo {author} {\bibfnamefont {J.~W.}\ \bibnamefont {Koo}}, \bibinfo {author} {\bibfnamefont {H.}~\bibnamefont {Sukegawa}}, \bibinfo {author} {\bibfnamefont {S.}~\bibnamefont {Mitani}}, \bibinfo {author} {\bibfnamefont {Y.}~\bibnamefont {Takagi}},\ and\ \bibinfo {author} {\bibfnamefont {T.}~\bibnamefont {Yokoyama}},\ }\href@noop {} {\bibfield  {journal} {\bibinfo  {journal} {Appl. Phys. Lett.}\ }\textbf {\bibinfo {volume} {105}},\ \bibinfo {pages} {122408} (\bibinfo {year} {2014})}\BibitemShut {NoStop}%
\bibitem [{\citenamefont {Okabayashi}\ \emph {et~al.}(2020)\citenamefont {Okabayashi}, \citenamefont {Miura}, \citenamefont {Kota}, \citenamefont {Z~Suzuki}, \citenamefont {Sakuma},\ and\ \citenamefont {Mizukami}}]{Okabayashi2020-dn}%
  \BibitemOpen
  \bibfield  {author} {\bibinfo {author} {\bibfnamefont {J.}~\bibnamefont {Okabayashi}}, \bibinfo {author} {\bibfnamefont {Y.}~\bibnamefont {Miura}}, \bibinfo {author} {\bibfnamefont {Y.}~\bibnamefont {Kota}}, \bibinfo {author} {\bibfnamefont {K.}~\bibnamefont {Z~Suzuki}}, \bibinfo {author} {\bibfnamefont {A.}~\bibnamefont {Sakuma}},\ and\ \bibinfo {author} {\bibfnamefont {S.}~\bibnamefont {Mizukami}},\ }\href@noop {} {\bibfield  {journal} {\bibinfo  {journal} {Sci. Rep.}\ }\textbf {\bibinfo {volume} {10}},\ \bibinfo {pages} {9744} (\bibinfo {year} {2020})}\BibitemShut {NoStop}%
\bibitem [{\citenamefont {Matsumura}\ \emph {et~al.}(2009)\citenamefont {Matsumura}, \citenamefont {Yonemura}, \citenamefont {Kunimori}, \citenamefont {Sera},\ and\ \citenamefont {Iga}}]{Matsumura2009-kg}%
  \BibitemOpen
  \bibfield  {author} {\bibinfo {author} {\bibfnamefont {T.}~\bibnamefont {Matsumura}}, \bibinfo {author} {\bibfnamefont {T.}~\bibnamefont {Yonemura}}, \bibinfo {author} {\bibfnamefont {K.}~\bibnamefont {Kunimori}}, \bibinfo {author} {\bibfnamefont {M.}~\bibnamefont {Sera}},\ and\ \bibinfo {author} {\bibfnamefont {F.}~\bibnamefont {Iga}},\ }\href@noop {} {\bibfield  {journal} {\bibinfo  {journal} {Phys. Rev. Lett.}\ }\textbf {\bibinfo {volume} {103}},\ \bibinfo {pages} {017203} (\bibinfo {year} {2009})}\BibitemShut {NoStop}%
\bibitem [{\citenamefont {Toyohiko}\ \emph {et~al.}(2012)\citenamefont {Toyohiko}, \citenamefont {Kuniaki}, \citenamefont {Keiki}, \citenamefont {Takuo}, \citenamefont {Masato}, \citenamefont {Fangzhun}, \citenamefont {Takayuki}, \citenamefont {Tetsuya}, \citenamefont {Hitoshi}, \citenamefont {Tomohiro},\ and\ \citenamefont {Taichi}}]{Toyohiko2012-ck}%
  \BibitemOpen
  \bibfield  {author} {\bibinfo {author} {\bibfnamefont {K.}~\bibnamefont {Toyohiko}}, \bibinfo {author} {\bibfnamefont {A.}~\bibnamefont {Kuniaki}}, \bibinfo {author} {\bibfnamefont {F.}~\bibnamefont {Keiki}}, \bibinfo {author} {\bibfnamefont {O.}~\bibnamefont {Takuo}}, \bibinfo {author} {\bibfnamefont {K.}~\bibnamefont {Masato}}, \bibinfo {author} {\bibfnamefont {G.}~\bibnamefont {Fangzhun}}, \bibinfo {author} {\bibfnamefont {M.}~\bibnamefont {Takayuki}}, \bibinfo {author} {\bibfnamefont {N.}~\bibnamefont {Tetsuya}}, \bibinfo {author} {\bibfnamefont {O.}~\bibnamefont {Hitoshi}}, \bibinfo {author} {\bibfnamefont {M.}~\bibnamefont {Tomohiro}},\ and\ \bibinfo {author} {\bibfnamefont {O.}~\bibnamefont {Taichi}},\ }\href@noop {} {\bibfield  {journal} {\bibinfo  {journal} {Journal of the Physical Society of Japan}\ } (\bibinfo {year} {2012})}\BibitemShut {NoStop}%
\bibitem [{\citenamefont {Higuchi}\ and\ \citenamefont {Kuwata-Gonokami}(2016)}]{higuchi2016control}%
  \BibitemOpen
  \bibfield  {author} {\bibinfo {author} {\bibfnamefont {T.}~\bibnamefont {Higuchi}}\ and\ \bibinfo {author} {\bibfnamefont {M.}~\bibnamefont {Kuwata-Gonokami}},\ }\href@noop {} {\bibfield  {journal} {\bibinfo  {journal} {Nature communications}\ }\textbf {\bibinfo {volume} {7}},\ \bibinfo {pages} {10720} (\bibinfo {year} {2016})}\BibitemShut {NoStop}%
\bibitem [{\citenamefont {Koizumi}\ \emph {et~al.}(2023)\citenamefont {Koizumi}, \citenamefont {Yamasaki},\ and\ \citenamefont {Yanagihara}}]{koizumi2023quadrupole}%
  \BibitemOpen
  \bibfield  {author} {\bibinfo {author} {\bibfnamefont {H.}~\bibnamefont {Koizumi}}, \bibinfo {author} {\bibfnamefont {Y.}~\bibnamefont {Yamasaki}},\ and\ \bibinfo {author} {\bibfnamefont {H.}~\bibnamefont {Yanagihara}},\ }\href@noop {} {\bibfield  {journal} {\bibinfo  {journal} {Nature Communications}\ }\textbf {\bibinfo {volume} {14}},\ \bibinfo {pages} {8074} (\bibinfo {year} {2023})}\BibitemShut {NoStop}%
\end{thebibliography}

\clearpage

\begin{table}[t]
    \begin{center}
    \begin{tabular}{|c|c|c|c|}
        \hline
        $(s,k)$ & $(1,-1)$ & $(1,0)$ & $(1,1)$ \\
        \hline
        $Q_{00}^{(s,k)}$  & - & - & $-C_{1,m-n;1n}^{00}M_{1,m-n}^{(\text{orb})}\sigma_{1n}$\\
        \hline
        $M_{1m}^{(s,k)}$  & 
        $C_{0,m-n;1n}^{1m} Q^{\text{(orb)}}_{0,m-n}\sigma_{1n}$  & 
        $iC_{1,m-n;1n}^{1m}G^{\text{(orb)}}_{1,m-n}\sigma_{1n}$ &
        $-C_{2,m-n;1n}^{1m}Q^{\text{(orb)}}_{2,m-n}\sigma_{1n}$ \\\hline
        \( Q_{2m}^{(s,k)} \)  & $C_{1,m-n;1n}^{2m}M^{\text{(orb)}}_{1,m-n}\sigma_{1n}$  & $iC_{2,m-n;1n}^{2m}T^{\text{(orb)}}_{2,m-n}\sigma_{1n}$ & $-C_{3,m-n;1n}^{2m}M^{\text{(orb)}}_{3,m-n}\sigma_{1n}$\\
        \hline
    \end{tabular}
    \vspace{5mm}
    \caption{Classification of spinful multiples ($s=1$) with the inversion symmetry, \textit{i.e.} electric monopole, magnetic dipole, and electric quadrupole moment, for possible $k$ parameters.}
    \label{tab:multipoles}
    \end{center}
\end{table}

\begin{table}[t]
    \centering
    \begin{tabular}{|c|c|}
\hline
$(l,m)$&$\nabla O_{lm}$\\
\hline
$(1,\pm 1)$&$\mp \frac{1}{2}(e_x \pm ie_y)$\\
$(1,0)$ & $e_z$\\
\hline
$(3,\pm 3)$ & $\mp \frac{3\sqrt{5}}{4} (x \pm iy)^2 (e_x \pm i e_y)$\\
$(3,\pm 2)$ & $\frac{1}{2} \sqrt{\frac{15}{2}} (x \pm iy) \left[ 2z (e_x \pm i e_y) + (x \pm iy) e_z \right]$\\
$(3,\pm 1)$ & $\mp\frac{\sqrt{3}}{4} \left[ (5z^2 - r^2)(e_x \pm i e_y) + 2(x \pm iy)(5z e_z - r) \right]$\\
$(3,0)$&$-3z(x e_x + y e_y) + \frac{3}{2} (3z^2 - r^2) e_z$\\
\hline
    \end{tabular}
    \vspace{5mm}
    \caption{Explicit expression of $\nabla O_{lm}$ relating magnetic dipole ($l=1$) and octupole ($l=3$) with the position operator $\bm{r}=(x,y,z)$ and the unit vector $\bm{e} =(e_x,e_y, e_z)$.}
    \label{nabla_O_lm}
\end{table}

\begin{figure}
\centering
\resizebox*{12cm}{!}{\includegraphics{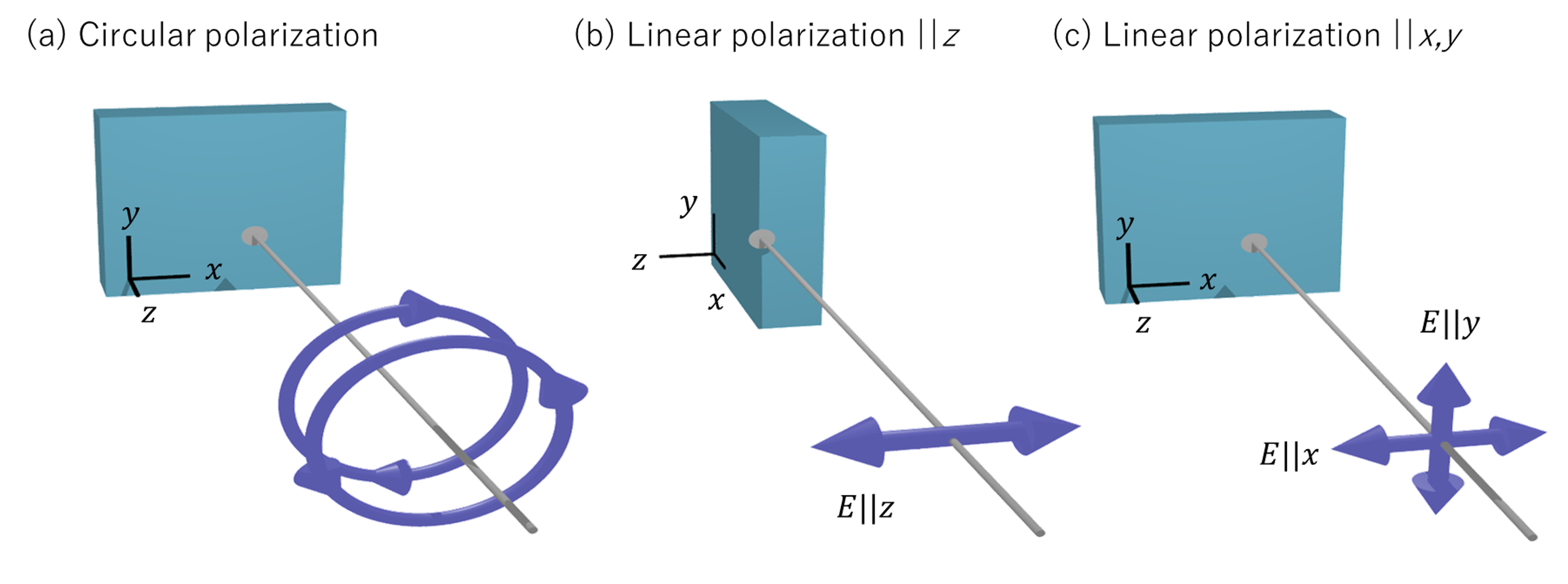}}
\caption{Experimental setup of x-ray absorption with the incident polarization of (a) circular, (b) linear ($E||z$), and (c) linear ($E||x$ and $E||y$).} \label{sample-figure}
\end{figure}

\begin{table*}[t]\label{technique_property}
\begin{center}
\begin{tabular}{|c|c|c|c|c|c|c|}
\hline
Technique &$l$& $m$ & $s$ & $k$ & multipole basis& sum rule\\
\hline\hline
{XAS} &{0} & 0 & 0 & 0 & $Q_{00}^{(orb)}\sigma_0$ & $n_h$\\
\cline{4-7}
 &&  & 1 & 1 & $M^{(orb)}_{1,-n}\sigma_{1n}$ & $\underline{\hat{\bm{l}}}\cdot \underline{\hat{\bm{s}}}$ \\
\hline
XMCD & 1 & $m$ & 0 & 0 & $M^{(orb)}_{1,m}\sigma_{0}$ & $\underline{l}_z$ \\
\cline{4-7}
& &  & 1 & -1 & $Q^{(orb)}_{0,0}\sigma_{1n}$ & $\underline{s}_z$ \\
& &  & 1 & 1 & $Q^{(orb)}_{2,m-n}\sigma_{1n}$ & $[\underline{\bm{Q}} \otimes \underline{\bm{s}}]_z$ \\
\hline
XMLD& 2 & $m$ & 0 & 0 & $Q^{(orb)}_{2,m}\sigma_{0}$ & $\underline{Q}_{3z^2-r^2}, \underline{Q}_{x^2-y^2}$ \\
\cline{4-7}
& &  & 1 & -1 & $M^{(orb)}_{1,m-n}\sigma_{1n}$ & $[\bm{\underline{l}} \otimes  \bm{\underline{s}}]_z$ \\
& &  & 1 & 1 & $M^{\text{(orb)}}_{3,m-n}\sigma_{1n}$ &  $[\bm{\underline{M}}^{\rm (orb)}_3\otimes  \bm{\underline{s}}]_z$\\
\hline
\end{tabular}
\vspace{5mm}
\caption{Classification of sum rules for X-ray absorption spectrum (XAS), X-ray magnetic circular dichroism (XMCD), and X-ray magnetic linear dichroism (XMLD) based on the complete multipole basis. }
\end{center}
\end{table*}

\end{document}